\documentclass[aps]{revtex4}
\usepackage{eurosym}
\usepackage{amsfonts}
\usepackage{amsmath}
\usepackage{amssymb,epsf}
\usepackage{color}
\usepackage{xcolor}
\usepackage{graphicx}
\usepackage{epstopdf}
\usepackage{float}
\usepackage{caption}
\usepackage{subfig}
\usepackage{comment}
\usepackage{ulem}
\usepackage{hyperref}
\usepackage{mathtools}

\begin{document}

\title{Accelerating electrically charged ModMax black hole solutions in $F(R)$ gravity}
\author{ B. Eslam Panah$^{1,2}$ \footnote{email address: eslampanah@umz.ac.ir/behzad.eslampanah@gmail.com}, N. Heidari 
$^{1}$, and {\'A}. Rinc{\'o}n $^{3,4}$ \footnote{%
email address: angel.rincon@physics.slu.cz } }
\affiliation{$^{1}$ Department of Theoretical Physics, Faculty of Basic Sciences,
University of Mazandaran, P. O. Box 47416-95447, Babolsar, Iran,\\
$^{2}$ Center for Theoretical Physics, Khazar University, 41 Mehseti Str., Baku, AZ1096, Azerbaijan,\\
$^{3}$ Departamento de F{\'i}sica, Universidad del B{\'i}o-B{\'i}o, Casilla 5-C, Concepci{\'o}n, Chile,\\
$^{4}$ Research Centre for Theoretical Physics and Astrophysics, Institute of Physics, Silesian University in Opava, Bezru\v{c}ovo~n\'am\v{e}st\'i 13, CZ-74601 Opava, Czech Republic.}

\begin{abstract}
Using the $C-$metric in the context of $F(R)$ gravity coupled with the
ModMax nonlinear electromagnetic field (the $F(R)-$ModMax theory), we derive
an exact black hole solution in a four-dimensional spacetime. We then
examine how various parameters influence the behavior of accelerating ModMax
black holes. Treating this black hole as a thermodynamic system, we
calculate the Hawking temperature and entropy for the accelerating ModMax
black holes within the framework of $F(R)$ gravity. Subsequently, we explore
the impact of the parameters in $F(R)-$ModMax theory on the Hawking
temperature and entropy. We also assess local stability by analyzing the
heat capacity. Additionally, we investigate both the angular shadow and the
shadow radius of an accelerating black hole in the context of $F(R)-$ModMax
gravity.
\end{abstract}

\maketitle

\section{\textbf{Introduction}}

Various observations emphasize the existence of the late-time accelerated
expansion of the Universe \cite{AccUn1,AccUn2,AccUn3,AccUn4}. This
accelerated expansion reveals a profound challenge to both the standard
cosmological model and General Relativity (GR), as GR cannot adequately
explain this phenomenon. Consequently, there has been considerable interest
in modified theories of gravity aimed at accounting for cosmic acceleration
without relying on a cosmological constant or exotic dark energy components.
Of the plethora of alternative theories of gravity, there are some concrete
scenarios where the modifications are minimal, maintaining the prediction of
GR untouched and mitigating/reducing the intrinsic problems of classical GR
(as the problem of singularities, the Hubble tension and the cosmological
constant problem). Special attention is devoted to a reduced number of them:
ranging from $F(R)$ gravity \cite{DeFelice:2010aj}, scalar-tensor theories 
\cite{Fujii:2003pa}, and loop quantum gravity \cite%
{Ashtekar:2021kfp,Rovelli:1997yv}, to teleparallel gravity \cite%
{Krssak:2018ywd}, Horava-Lifshitz gravity \cite{Wang:2017brl} and
quantum-inspired theories of gravity \cite%
{Bonanno:2000ep,Koch:2016uso,Rincon:2018sgd,Eichhorn:2020mte}. Among these, $%
F(R)$ gravity stands out as one of the most popular and viable alternatives 
\cite{FRI,FRII,FRIII,FRIV,FRV}. $F(R)$ gravity is formulated by generalizing
the Einstein-Hilbert action to include an arbitrary function of the Ricci
scalar $R$. This theory belongs to a class of models that incorporate higher-order curvature invariants, providing a robust geometric explanation for the observed cosmic acceleration \cite{FRI,FRII,FRIII,FRIV,FRV}. $F(R)$ gravity has been rigorously investigated across various cosmological and astrophysical contexts. For instance, it provides potential resolutions to the Hubble tension \cite{HubTen} and effectively describes the inflationary phase of the early Universe \cite{Starobinsky,Starobinsky2,Inflation1,Inflation2,Inflation3,Inflation4,Inflation5}. A prominent example is the Starobinsky inflationary model, originally proposed in \cite{Starobinsky}, which is based on an $R+\alpha R^{2}$ gravity. This framework has been extended to various Starobinsky-like models, such as the power-law form $F(R)=R+\alpha R^{n}$ \cite{Starobinsky2}. In this formulation, $n=2$ corresponds to the original Starobinsky model, while cases with $n>2$ result in super-inflation. Another common parametrization is $F(R)=R+\frac{R^{2}}{6M^{2}}$ \cite{Starobinsky2}. Furthermore, $F(R)$ gravity addresses dark matter phenomenology \cite{DM1,DM2,DM3,DM4,DM5} and remains consistent with both Newtonian and post-Newtonian approximations \cite{Newton1,Newton2}. Additionally, this theory predicts the existence of massive compact objects \cite{massiveCom1,massiveCom2,massiveCom3,massiveCom4,massiveCom5,massiveCom6,massiveCom7,massiveCom8,massiveCom9} and aligns with rigorous solar system constraints \cite{solar1,solar2,solar3,solar4}.

In theoretical physics, black hole solutions arise from the field equations
of gravitational theories. These solutions typically feature a singularity
and an event horizon. However, some black hole solutions, known as regular
black holes, do not contain singularities. The exploration of black hole
solutions has greatly enhanced our understanding of spacetime singularities,
event horizons, and causal structures. While the concept of black holes is
rooted in theoretical physics, the first evidence of their existence was
reported by the Advanced LIGO/Virgo collaboration through gravitational
waves in 2016 \cite{LIGO}. Subsequently, in 2019, the Event Horizon
Telescope captured the first image of a black hole \cite{ETH}. These
remarkable discoveries have inspired many to study black hole physics in
order to explore constraints on the parameters of modified theories.

The presence of singularities in physics is always a challenging topic. For
instance, in classical electromagnetic theory (Maxwell's theory), a
singularity arises at the location of a point-like charge. To address this
issue, nonlinear electrodynamics (NLED) theories, such as Born$-$Infeld
theory \cite{BI}, have been developed. Born$-$Infeld theory originates from
open superstring theory and offers a non-singular self-energy for point-like
charges \cite{BI1,BI2}. Furthermore, NLED theories can eliminate
singularities related to the Big Bang and black holes \cite%
{NED1,NED2,NED3,NED4,NED5,NED6,NED7}. It is worth noting that recent
studies \cite{CausalNED1,CausalNED2} have shown that regular black hole
solutions can arise only in NLED theories that violate causality. In
contrast, causal NLED theories, such as ModMax electrodynamics, admit only
black hole solutions with singularities. NLED has also been shown to
influence gravitational redshift around strongly magnetic compact objects \cite{NEDMag1,NEDMag2}. Considering the significance of black hole solutions in $F(R)$ gravity and NLED fields, black hole solutions in $F(R)$ gravity coupled to NLED have been obtained in Refs. \cite%
{FRNLEDBH1,FRNLEDBH2,FRNLEDBH3,FRNLEDBH4,FRNLEDBH5,FRNLEDBH6,FRNLEDBH7,FRNLEDBH8}. It is important to note that Maxwell's theory incorporates two
fundamental symmetries: electromagnetic duality and conformal invariance.
The existence of NLED theories that preserve these symmetries is crucial.
Unfortunately, Born$-$Infeld theory fails in this regard, as it does not
maintain conformal invariance \cite{BI3}. To address this limitation, a new
model of NLED, known as Modification of Maxwell (ModMax) theory, was
proposed in 2020, which successfully retains both symmetries \cite%
{ModMaxI,ModMaxII}. This new NLED theory introduces a dimensionless
parameter $\gamma$, referred to as the ModMax parameter. The ModMax NLED
theory reduces to Maxwell's theory when the ModMax parameter is zero, i.e., $%
\gamma=0$. Research has also examined the effects of the ModMax NLED field
on various properties of black holes, including shadow \cite{ShadowModMax},
light propagation \cite{LightModMax}, emission rates \cite{quasiModMax},
gravitational lensing \cite{GraLensModMax}, quasinormal modes \cite%
{ShadowModMax,quasiModMax}, greybody radiation \cite{ShadowModMax}, and
thermodynamic topology \cite{TheTopoModMax}. Additionally, further
investigations into the intriguing aspects of ModMax theory are presented in
Refs. \cite%
{ModMaxMagnetic,ModMaxElecMag,MM1,MM2,MM3,MM4,MM5,MM6,MM7,MM8,MM9,MM10,MM11,MM12,MM13,MM14,MM15,MM16,MM17,MM18,MM19,MM20,MM21,MM22,MM23}%
. Within the context of ModMax NLED theory, various black hole solutions
have been derived, including accelerating \cite%
{acc1ModMax,acc2ModMax,acc3ModMax} and BTZ \cite{BTZModMax} black holes.

On the other hand, the black hole solutions derived from the $C-$metric are
classified as accelerating black holes \cite%
{Kinnersley,Plebanski,Dias,Griffiths}. These black holes exhibit distinct
behaviors compared to other well-known black hole solutions, primarily due
to the characteristics of the $C-$metric. A notable distinction is the
presence of a conical deficit angle along one polar axis, which generates
the force responsible for their acceleration. Furthermore, the $C-$metric
has important applications within the context of AdS/CFT correspondence. For
instance, it enables the construction of exact black holes on branes \cite%
{Emparan2000}, plasma balls \cite{Emparan2009}, black funnels and droplets 
\cite{Hubeny2010}, and spinning spindles derived from accelerating black
holes \cite{Ferrero2021}. Additionally, accelerated black holes have been
demonstrated to possess a higher Hawking temperature than the Unruh
temperature associated with their accelerated frame \cite{Letelier}. Their
asymptotic behaviors are complex and vary based on different parameters. A
significant study established the first law of black hole thermodynamics for
charged accelerating black holes, adhering to the conventional
identification of entropy as proportional to the area of the event horizon 
\cite{The1}. The thermodynamics of accelerating black holes in AdS spacetime
was investigated in \cite{The4}, revealing that both holographic
computations and conformal completion methods yield consistent results for
the mass of these black holes. The thermodynamic properties of charged
rotating accelerating black holes were examined in \cite{Anabalon2019,The7},
and a new set of chemical variables for these black holes was introduced in 
\cite{The5}. This research proposed that conical defects associated with a
black hole could be regarded as legitimate hair and discussed the
implications of these conical deficits on black hole thermodynamics from a
'chemical' perspective. Consequently, the exploration of the thermodynamic
properties of accelerating AdS black holes in (non-)extended phase space has
attracted considerable interest (see, for example, \cite%
{The2,The8a,The8b,The8c,The8}). Furthermore, accelerating black holes with a
conformally coupled scalar field in a magnetic universe were analyzed in 
\cite{Astorino2013}, while accelerating Kerr-Newman black holes were studied
in \cite{Astorino2016}. It was demonstrated that, at extremality, these
black holes remain a warped and twisted product of AdS. Additionally, the
applicability of the Kerr/CFT correspondence to the accelerating and
magnetized extremal black holes was noted. When interpreting the $C-$metric
as a gravitational field representing an accelerating black hole influenced
by a semi-infinite cosmic string (in the direction of acceleration), the
gravitational energy enclosed by surfaces of constant radius around the
black hole was assessed. The results indicated that the gravitational energy
of the semi-infinite cosmic string is negative, which can explain the
acceleration of the black hole as it moves toward regions of lower
gravitational energy along the string \cite{Carneiro2022}.

The combination of ModMax with various modified theories of gravity can
facilitate the exploration of new black hole solutions and their properties.
For instance, black holes in massive$-$ModMax theory \cite{massiveModMax},
Kalb$-$Ramond$-$ModMax theory \cite{KR1ModMax, KR2ModMax, KR3ModMax}, and $%
F(R)-$ModMax theory \cite{FRModMax} have been studied by examining static,
spherically symmetric spacetimes. Accelerating black holes in $F(R)-$Maxwell
gravity have been evaluated in Ref. \cite{FRAcele}. However, there is
currently no solution for accelerating black holes within the
framework of $F(R)$ gravity with NLED field such as $F(R)-$ModMax theory.
Our objective is to derive an exact accelerating black hole solution in $%
F(R)-$ModMax theory. Subsequently, we will investigate various thermodynamic
and optical properties of these black holes.

The work is organized as follows: after this detailed introduction, we will
introduce the action in $F(R)$ gravity in the presence of NLED, particularly
in the ModMax case, as well as the corresponding equation of motion. Section
III provides a brief summary of the $C-$metric and the accelerating black
hole solutions found here. It shows not only the solutions, but also the
Kretschmann scalar and the influence of several of the parameters involved.
Section IV then analyses the thermal stability and some thermodynamic
properties in detail, both analytically and graphically. Section V then
investigates the optical appearance, including the photon sphere radius,
celestial coordinates, and angular shadow and shadow radius. The final
section provides a summary of our main findings.

Briefly, this work extends the research on accelerating ModMax black holes
in general relativity \cite{acc2ModMax} and the accelerating charged black
holes within $F(R)$ gravity \cite{FRAcele}. Specifically, we extract the
first accelerating black hole solutions in the presence of a nonlinear
electrodynamics field (the ModMax field) under the framework of $F(R)$
gravity.

\section{Action and field equations}

The action of $F(R)$ gravity in the presence of ModMax NLED is given by 
\begin{equation}
I=\frac{1}{16\pi }\int d^{4}x\sqrt{-g}[F(R)-4\mathcal{L}],  \label{action}
\end{equation}%
here, we consider $G=c=1$ in the action (\ref{action}), where $c$ and $G$
are the speed of light and the gravitational constant, respectively. In
addition, $g$ devotes to the determinant of metric tensor $g_{\mu \nu }$,
i.e., $g=det(g_{\mu \nu })$. Also, $F(R)=R+f\left( R\right) $, where $f\left( R\right) $ is a function of the Ricci
scalar ($R$). In the action (\ref{action}), $\mathcal{L}$ is
related to the ModMax's Lagrangian and is defined by \cite{ModMaxI,ModMaxII,MM7} 
\begin{equation}
\mathcal{L}=\frac{-1}{2}\left( \mathcal{S}\cosh \gamma -\sqrt{\mathcal{S}%
^{2}+\mathcal{P}^{2}}\sinh \gamma \right) ,  \label{ModMaxL}
\end{equation}%
where $\mathcal{S}=\frac{\mathcal{F}}{2}$ and $\mathcal{P}=\frac{\widetilde{%
\mathcal{F}}}{2}$, respectively, are a true scalar, and a pseudoscalar. $%
\gamma $ and $\mathcal{F}=F_{\mu \nu }F^{\mu \nu }$ are, respectively, a
dimensionless parameter (which is known as the ModMax's parameter), and the
Maxwell invariant. It is notable that, $F_{\mu \nu }=\partial _{\mu }A_{\nu
}-\partial _{\nu }A_{\mu }$ is the electromagnetic tensor field (where $%
A_{\mu }$ is the gauge potential). Also, $\widetilde{\mathcal{F}}=$ $F_{\mu
\nu }\widetilde{F}^{\mu \nu }$, and $\widetilde{F}^{\mu \nu }=\frac{1}{2}%
\epsilon _{\mu \nu }^{~~~\rho \tau }F_{\rho \tau }$. Notably, the ModMax Lagrangian (Eq. (\ref{ModMaxL})) reduces to the standard Maxwell theory in the limit $\gamma \longrightarrow 0$. Since the present work focuses on deriving purely electrically charged black hole solutions, we consider the vanishing magnetic parameter ($\mathcal{P}=0$) in Eq. (\ref{ModMaxL}). In this regime, the ModMax Lagrangian effectively takes a Maxwell-like form scaled by the factor $e^{\gamma}$ \cite{MM14}. It should be emphasized that while our current solutions explore the electric sector, the broader implications of ModMax non-linearity, including the role of magnetic charges, remain essential considerations for a comprehensive understanding of the theory.

In this work, we are interested in considering the electrically charged case
of the ModMax theory. In other words, we want to obtain electrical charged
black hole solutions by coupling $F(R)$ theory and the ModMax NLED theory.
Therefore, we have to consider $\mathcal{P}=0$ in the above equations. For
this purpose, we are able to extract the equations of motion of $F(R)-$%
ModMax theory of gravity, which lead to 
\begin{eqnarray}
R_{\mu \nu }\left( 1+f_{R}\right) -\frac{g_{\mu \nu }F(R)}{2}+\left( g_{\mu
\nu }\nabla ^{2}-\nabla _{\mu }\nabla _{\nu }\right) f_{R} &=&8\pi \mathrm{T}%
_{\mu \nu },  \label{EqF(R)1} \\
&&  \notag \\
\partial _{\mu }\left( \sqrt{-g}\widetilde{E}^{\mu \nu }\right) &=&0,
\label{EqF(R)2}
\end{eqnarray}%
where $f_{R}=\frac{df(R)}{dR}$. Also, $\mathrm{T}_{\mu \nu }$ defines as the
energy-momentum tensor, which is given by 
\begin{equation}
4\pi \mathrm{T}^{\mu \nu }=\left( F^{\mu \sigma }F_{~~\sigma }^{\nu
}e^{-\gamma }\right) -\frac{\mathcal{S}}{2}e^{-\gamma }g^{\mu \nu },
\label{eq3}
\end{equation}%
and $\widetilde{E}_{\mu \nu }$ in Eq. (\ref{EqF(R)2}), is defined as 
\begin{equation}
\widetilde{E}_{\mu \nu }=\frac{\partial \mathcal{L}}{\partial F^{\mu \nu }}%
=2\left( \mathcal{L}_{\mathcal{S}}F_{\mu \nu }\right) ,  \label{eq3b}
\end{equation}%
where $\mathcal{L}_{\mathcal{S}}=\frac{\partial \mathcal{L}}{\partial 
\mathcal{S}}$. So, the ModMax field equation (Eq. (\ref{EqF(R)2})) for the
electrically charged case reduces to 
\begin{equation}
\partial _{\mu }\left( \sqrt{-g}e^{-\gamma }F^{\mu \nu }\right) =0.
\label{Maxwell Equation}
\end{equation}

\section{$C-$metric and Accelerating Black Hole Solutions}

In order to extract the accelerating ModMax black hole in $F(R)$ gravity, we
consider $C-$metric with metric signature $(-,+,+,+)$ in the following form 
\begin{equation}
ds^{2}=\frac{1}{\mathcal{K}^{2}\left( r,\theta \right) }\left[ -g(r)dt^{2}+%
\frac{dr^{2}}{g(r)}+r^{2}\left( \frac{d\theta ^{2}}{X\left( \theta \right) }+%
\frac{X\left( \theta \right) \sin ^{2}\theta d\varphi ^{2}}{K^{2}}\right) %
\right] ,  \label{Metric}
\end{equation}%
where $\mathcal{K}\left( r,\theta \right) =1+Ar\cos \theta $, which is
called the conformal factor. Notably, we can define $K$ as introduced in
Refs. \cite{BHacc1,BHacc4}, which is related to the presence of cosmic
string. In other words, by looking at the angular part of the metric and the
behavior of $X\left( \theta \right) $ at both poles $\theta _{+}=0$ (north
pole), and $\theta _{-}=\pi $ (south pole), we can find the presence of
cosmic string. The regularity of the metric at a pole requires $K_{\pm
}=X\left( \theta _{\pm }\right) =1\pm 2m_{0}A$, where $K_{\pm }$ is chosen
to regularize one pole and another pole is left with either a conical
deficit or a conical excess along the other pole. It is notable that, by
considering $\mathcal{K}\left( r,\theta \right) =1$, $X\left( \theta \right)
=1$ and $K=1$, the $C-$metric (\ref{Metric}) turns to the spherical
symmetric spacetime.

The equations governing $F(R)$ gravity with a nonlinear matter field, as
presented in Eq. (\ref{EqF(R)1}), are quite complex, making it difficult to
derive a precise analytical solution. To tackle this issue, one can utilize
the traceless energy-momentum tensor for the nonlinear matter field, such as
the ModMax field. This approach allows for the extraction of an exact
analytical solution from $F(R)$ gravity coupled with a nonlinear matter
field. To find the solution for a black hole with constant curvature in $%
F(R) $ theory of gravity in conjunction with the ModMax field, it is
essential that the trace of the stress-energy tensor $\mathrm{T}_{\mu \nu }$
equals zero \cite{Rcont1,Rcont2}. Assuming a constant scalar curvature,
specifically $R=R_{0}$, the trace of the equation (\ref{EqF(R)1}) simplifies
to 
\begin{equation}
R_{0}\left( 1+f_{R_{0}}\right) -2\left( R_{0}+f(R_{0})\right) =0,
\label{R00}
\end{equation}%
where $f_{R_{0}}=$ $f_{R_{\left\vert _{R=R_{0}}\right. }}$. We can solve the
equation (\ref{R00}) to obtain $R_{0}$ which leads to 
\begin{equation}
R_{0}=\frac{2f(R_{0})}{f_{R_{0}}-1}.  \label{R0}
\end{equation}

By replacing Eq. (\ref{R0}) within Eq. (\ref{EqF(R)1}), the $F(R)$-ModMax
theory's equations of motion can be found in the following format 
\begin{equation}
R_{\mu \nu }\left( 1+f_{R_{0}}\right) -\frac{g_{\mu \nu }}{4}R_{0}\left(
1+f_{R_{0}}\right) =8\pi \mathrm{T}_{\mu \nu }.  \label{F(R)Trace}
\end{equation}

It is notable that, the equation of motion in $F(R)-$ModMax theory (\ref%
{F(R)Trace}) reduces to GR$-$ModMad theory of graviry when $f_{R_{0}}=0$.

To obtain a radial electric field, we take the following form for the gauge
potential 
\begin{equation}
A_{\mu }=h\left( r\right) \delta _{\mu }^{t},
\end{equation}%
By utilizing the provided gauge potential and equations (\ref{Maxwell
Equation}) and (\ref{Metric}), we are able to derive the subsequent
differential equation. 
\begin{equation}
\mathcal{K}^{4}\left( r,\theta \right) \left( rh^{\prime \prime
}(r)+2h^{\prime }(r)\right) =0,  \label{hh}
\end{equation}%
where, respectively, the prime and double prime represent the first and
second derivatives of $r$. The solution of the equation (\ref{hh}) yields 
\begin{equation}
h(r)=-\frac{q}{r},  \label{h(r)}
\end{equation}%
where $q$ represents an integration constant that is associated with the
electric charge.

We are now able to obtain precise analytical solutions for the metric functions $g\left( r\right) $ and $X\left( \theta \right) $ by taking into
account the metric (\ref{Metric}), the derived $h(r)$, and the field
equations (\ref{F(R)Trace}). Consequently, we derive the subsequent set of
differential equations 
\begin{eqnarray}
eq_{tt} &=&eq_{rr}=r^{2}\mathcal{K}^{2}\left( r,\theta \right) g^{\prime
\prime }(r)+2Ar\sin \theta \mathcal{K}\left( r,\theta \right) X^{\theta
}(\theta )+2r\left( 2-\mathcal{K}\left( r,\theta \right) \right) \mathcal{K}%
\left( r,\theta \right) g^{\prime }(r)  \notag \\
&&  \notag \\
&&+2r\left[ A\left( g\left( r\right) -X\left( \theta \right) \right) \cos
\theta \left( \mathcal{K}\left( r,\theta \right) -3\right) +3\left(
A^{2}X\left( \theta \right) +\frac{R_{0}}{12}\right) r\right] -\frac{%
2q^{2}e^{-\gamma }\mathcal{K}^{4}\left( r,\theta \right) }{r^{2}\left(
1+f_{R_{0}}\right) },  \label{eq11} \\
&&  \notag \\
eq_{\theta \theta } &=&eq_{\varphi \varphi }=\mathcal{K}^{2}\left( r,\theta
\right) X^{\theta \theta }(\theta )-\frac{\mathcal{K}\left( r,\theta \right) 
}{\sin \theta }\left( \cos \theta \left( \mathcal{K}\left( r,\theta \right)
-4\right) -4Ar\right) X^{\theta }(\theta )+2r\mathcal{K}\left( r,\theta
\right) g^{\prime }(r)  \notag \\
&&  \notag \\
&&-8Ar\left( g\left( r\right) -X\left( \theta \right) \right) \cos \theta
\left( 3A^{2}r^{2}-1\right) X\left( \theta \right) +\frac{r^{2}R_{0}}{2}%
+2g\left( r\right) +\frac{2q^{2}e^{-\gamma }\mathcal{K}^{4}\left( r,\theta
\right) }{r^{2}\left( 1+f_{R_{0}}\right) },  \label{eq22}
\end{eqnarray}%
where $X^{\theta }(\theta )=\frac{dX(\theta)}{d\theta }$ and $X^{\theta
\theta }(\theta )=\frac{d^{2}X(\theta)}{d\theta ^{2}}$. In addition, $%
eq_{tt} $, $eq_{rr}$, $eq_{\theta \theta }$ and $eq_{\varphi \varphi }$,
respectively, are the components of $tt$, $rr$, $\theta \theta $ and $%
\varphi \varphi $ of field equations (\ref{F(R)Trace}).

By utilizing the aforementioned differential equations, we can obtain an
axact solution for the constant scalar curvature ($R=R_{0}$= constant).
After careful consideration and performing several calculations, the metric
functions can be expressed in the subsequent forms 
\begin{eqnarray}
g(r) &=&\left( 1-A^{2}r^{2}\right) \left( 1-\frac{2m_{0}}{r}+\frac{%
q^{2}e^{-\gamma }}{\left( 1+f_{R_{0}}\right) r^{2}}\right) -\frac{R_{0}r^{2}%
}{12},  \label{g(r)F(R)} \\
&&  \notag \\
X\left( \theta \right) &=&1+2m_{0}A\cos \theta +\frac{q^{2}e^{-\gamma
}A^{2}\cos ^{2}\theta }{\left( 1+f_{R_{0}}\right) },  \label{X(th)F(R)}
\end{eqnarray}%
where $m_{0}$ is an integration constant. It is noteworthy that this
constant of integration is connected to the black hole's geometric mass.
Furthermore, any of the field equations (\ref{F(R)Trace}) is satisfied by
the obtained solutions (\ref{g(r)F(R)}) and (\ref{X(th)F(R)}). We should
limit ourselves to $f_{R_{0}}\neq -1$ in order to have physical solutions.
In addition, we can see the effect of ModMax's theory and $F(R)$ gravity on
the obtained solutions (Eqs. (\ref{g(r)F(R)}) and (\ref{X(th)F(R)})). In the
absence of $F(R)$ gravity (i.e., $f_{R_{0}}=0$), the solutions reduce to the
accelerating ModMax black holes in GR, which are \cite{acc2ModMax} 
\begin{eqnarray}
g(r) &=&\left( 1-A^{2}r^{2}\right) \left( 1-\frac{2m_{0}}{r}+\frac{%
q^{2}e^{-\gamma }}{r^{2}}\right) -\frac{R_{0}r^{2}}{12}, \\
&&  \notag \\
X\left( \theta \right) &=&1+2m_{0}A\cos \theta +q^{2}e^{-\gamma }A^{2}\cos
^{2}\theta .
\end{eqnarray}

The obtained solutions (\ref{g(r)F(R)}) and (\ref{X(th)F(R)}) reduces to
accelerating black holes in $F(R)$ gravity when $\gamma =0$ \cite{FRAcele} 
\begin{eqnarray}
g(r) &=&\left( 1-A^{2}r^{2}\right) \left( 1-\frac{2m_{0}}{r}+\frac{q^{2}}{%
\left( 1+f_{R_{0}}\right) r^{2}}\right) -\frac{R_{0}r^{2}}{12}, \\
&&  \notag \\
X\left( \theta \right) &=&1+2m_{0}A\cos \theta +\frac{q^{2}A^{2}\cos
^{2}\theta }{\left( 1+f_{R_{0}}\right) },
\end{eqnarray}

Ia addition, Reissner-Nordstr\"{o}m-(A)dS black hole is covered by
considering $f_{R_{0}}=0$, $R_{0}=4\Lambda $, $\gamma =0$, and $A=0$ i.e., 
\begin{equation}
g(r)=1-\frac{2m_{0}}{r}+\frac{q^{2}}{r^{2}}-\frac{\Lambda r^{2}}{3}.
\end{equation}

Here, we study the Kretschmann scalar ($R_{\alpha \beta \gamma \delta
}R^{\alpha \beta \gamma \delta }$) in order to find the singularity of
spacetime. Indeed, this quantity gives us information about the existence of
the singularity in spacetime. For this purpose, we calculate the Kretschmann
scalar of the spacetime (\ref{Metric}), which is 
\begin{equation}
R_{\alpha \beta \gamma \delta }R^{\alpha \beta \gamma \delta }=g^{\prime
\prime ^{2}}(r)+\frac{4g^{\prime ^{2}}(r)}{r^{2}}+\frac{\left( 2g\left(
r\right) +X^{\theta \theta }(\theta )-2X(\theta )+3\cot \theta X^{\theta
}(\theta )\right) ^{2}}{r^{4}},  \label{K}
\end{equation}%
and by replacing the metric function (\ref{g(r)F(R)}) within Eq. (\ref{K}),
we have 
\begin{eqnarray}
R_{\alpha \beta \gamma \delta }R^{\alpha \beta \gamma \delta } &=&\frac{%
\left( R_{0}+12A^{2}\right) ^{2}}{6}-\frac{4m_{0}A^{2}\left(
R_{0}+12A^{2}\right) }{r}+\frac{4\left( R_{0}+12A^{2}\right) B_{1}}{r^{2}}+%
\frac{32m_{0}^{2}A^{4}}{r^{2}}  \notag \\
&&  \notag \\
&&-\frac{96m_{0}A^{2}B_{1}}{r^{3}}+\frac{144B_{1}^{2}}{r^{4}}+\frac{%
16m_{0}B_{2}}{r^{5}}-\frac{48B_{3}}{r^{6}}-\frac{96m_{0}q^{2}e^{-\gamma }}{%
\left( 1+f_{R_{0}}\right) r^{7}}+\frac{56q^{4}e^{-2\gamma }}{\left(
1+f_{R_{0}}\right) ^{2}r^{8}},
\end{eqnarray}%
where $B_{1}$, $B_{2}$, and $B_{3}$ are 
\begin{eqnarray}
B_{1} &=&\frac{q^{2}e^{-\gamma }A^{2}\cos ^{2}\theta }{\left(
1+f_{R_{0}}\right) }+m_{0}A\cos \theta , \\
&&  \notag \\
B_{2} &=&\frac{q^{2}e^{-\gamma }A^{2}\left( 6\cos ^{2}\theta -1\right) }{%
\left( 1+f_{R_{0}}\right) }+6m_{0}A\cos \theta , \\
&&  \notag \\
B_{3} &=&\frac{q^{4}e^{-2\gamma }A^{2}\cos ^{2}\theta }{\left(
1+f_{R_{0}}\right) ^{2}}+\frac{q^{2}e^{-\gamma }m_{0}A\cos \theta }{\left(
1+f_{R_{0}}\right) }-m_{0}^{2},
\end{eqnarray}%
so, the obtained Kretschmann scalar includes three important points, which
are

i) It diverges at $r=0$, i.e., 
\begin{equation}
\lim_{r\longrightarrow 0}R_{\alpha \beta \gamma \delta }R^{\alpha \beta
\gamma \delta }\longrightarrow \infty ,
\end{equation}
there exists a curvature singularity situated at the coordinate $r=0$.

ii) The Kretschmann scalar is finite for $r\neq 0$.

iii) The effect of ModMax's parameter appears in the Kretschmann scalar.
Although, the divergence of the electrical field is removed by considering $%
\gamma \longrightarrow \infty $, this limit cannot remove the curvature
singularity at $r=0$. In other words, in the limit $\gamma \longrightarrow
\infty $, we have $\lim_{r\rightarrow 0}R_{\alpha \beta \gamma
\delta}R^{\alpha \beta \gamma \delta }\longrightarrow \infty $.

The asymptotical behavior of the Kretschmann scalar is given by 
\begin{equation}
\lim_{r\longrightarrow \infty }R_{\alpha \beta \gamma \delta }R^{\alpha
\beta \gamma \delta }\longrightarrow \frac{\left( R_{0}+12A^{2}\right) ^{2}}{%
6},  \notag
\end{equation}%
where indicate that the spacetime will not be asymptotically (A)dS. It is
notable that, the asymptotical behavior is independent of $\gamma $. In
other words, the parameter of ModMax does not affect the asymptotical
behavior of the spacetime.

Our objective is to identify the real roots of the acquired metric function (%
\ref{g(r)F(R)}) since these roots provide insights into the horizons (both
inner and outer) of the solution. Black holes are characterized by a
curvature singularity located at $r=0$, which is hidden behind at least one
event horizon. Interestingly, it is also possible to have black holes
without an event horizon, known as naked singularities.

\begin{figure}[tbph]
\centering
\includegraphics[width=0.35\linewidth]{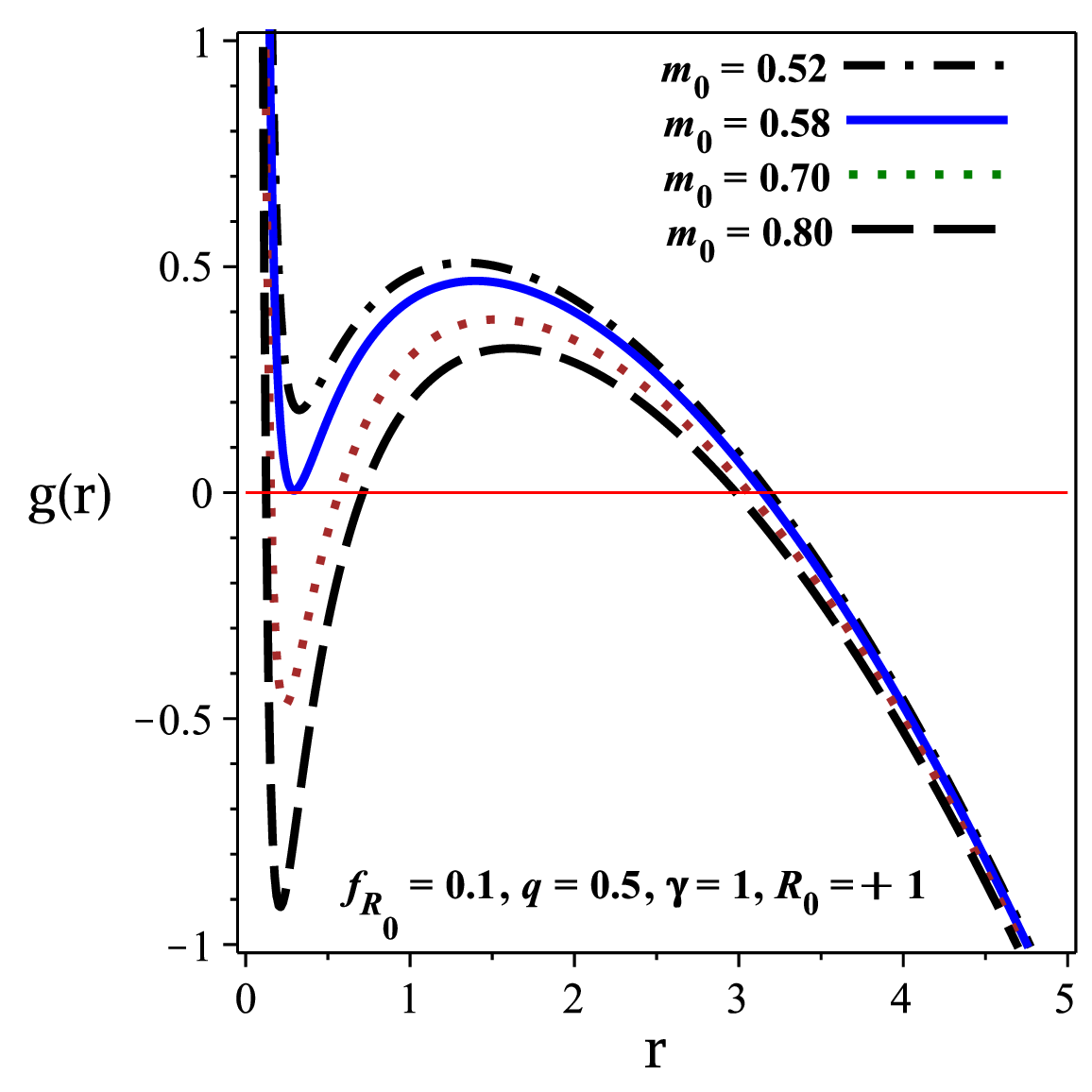} \includegraphics[width=0.35	%
\linewidth]{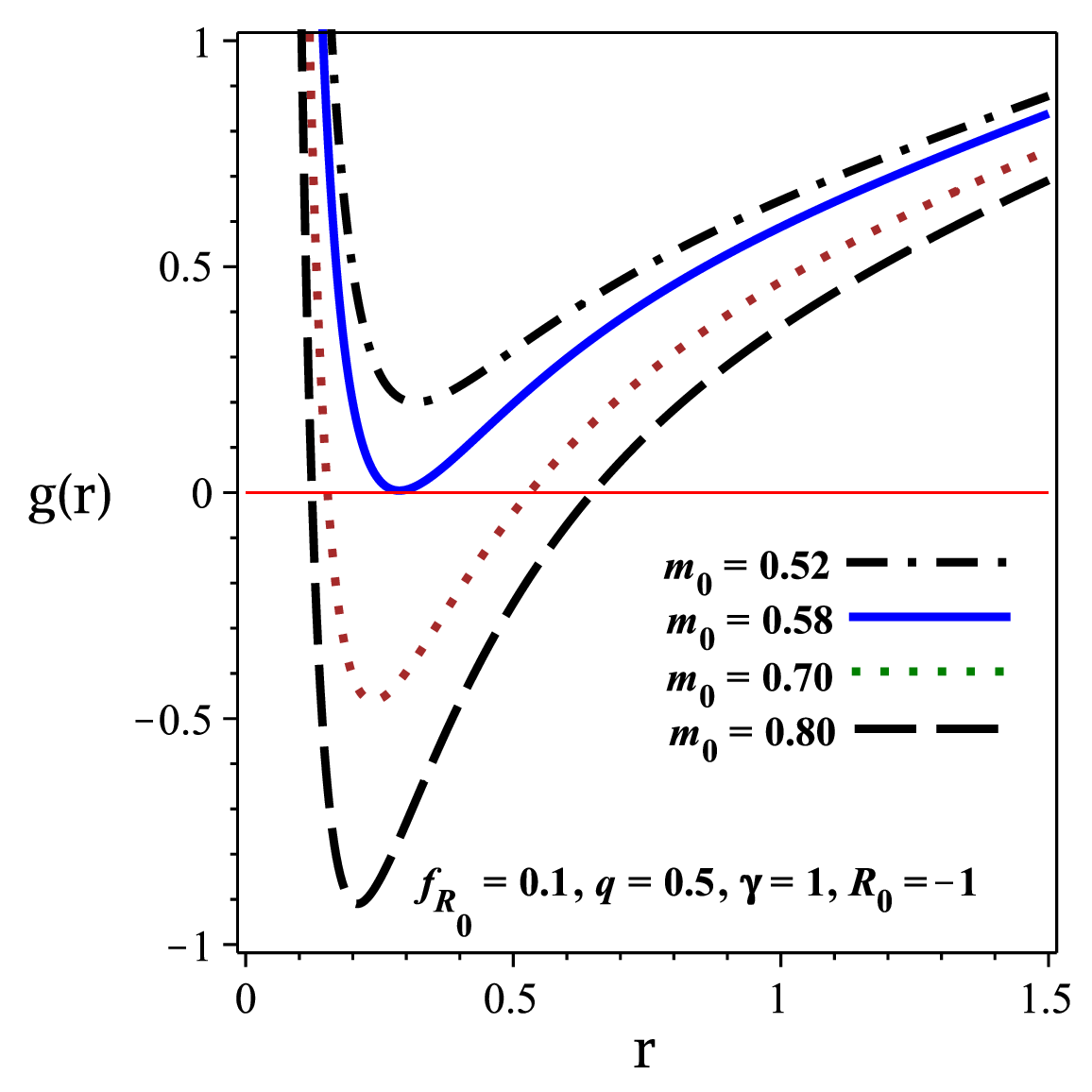}\newline
\caption{The function $g(r)$ is plotted against $r$ for various parameter
values, with the left panels corresponding to $R_{0}=1$ and the right panels
corresponding to $R_{0}=-1$.}
\label{Fig1}
\end{figure}

To find the roots, it is more effective to solve the metric function. The
metric function is a fourth-order function of $r$, making it challenging to
derive an exact solution. Consequently, we plot the metric function against
the variable $r$ in Fig. \ref{Fig1} to identify these roots. Our analysis
reveals the following:

i) For $R_{0} > 0$ (or the de Sitter case if we set $R_{0} = 4\Lambda$), as
shown in the left panel of Fig. \ref{Fig1}, we observe three different
scenarios based on the mass of the black holes. In the case of massive black
holes, there are three roots: an inner root, an event horizon, and an outer
root (the cosmological horizon). For black holes with medium mass, there are
two roots: an extreme case and the cosmological horizon. In contrast, for
black holes with lower mass, only one root (the cosmological horizon) exists.

ii) For $R_{0} < 0$ (or the anti-de Sitter case if we consider $R_{0} =
4\Lambda$), the solution can exhibit three different scenarios: i) two roots
(an inner horizon and an event horizon) for massive black holes; ii) one
root (an extreme case) for medium mass black holes; iii) a naked singularity
for black holes with lower mass. By adjusting certain parameters, it is
possible to create an event horizon that encloses the singularity located at 
$r=0$. These results confirm that the solution obtained in Eq. (\ref%
{g(r)F(R)}) may be linked to the accelerating black hole solution in $F(R)-$%
ModMax theory.

\begin{figure}[tbph]
\centering
\includegraphics[width=0.35\linewidth]{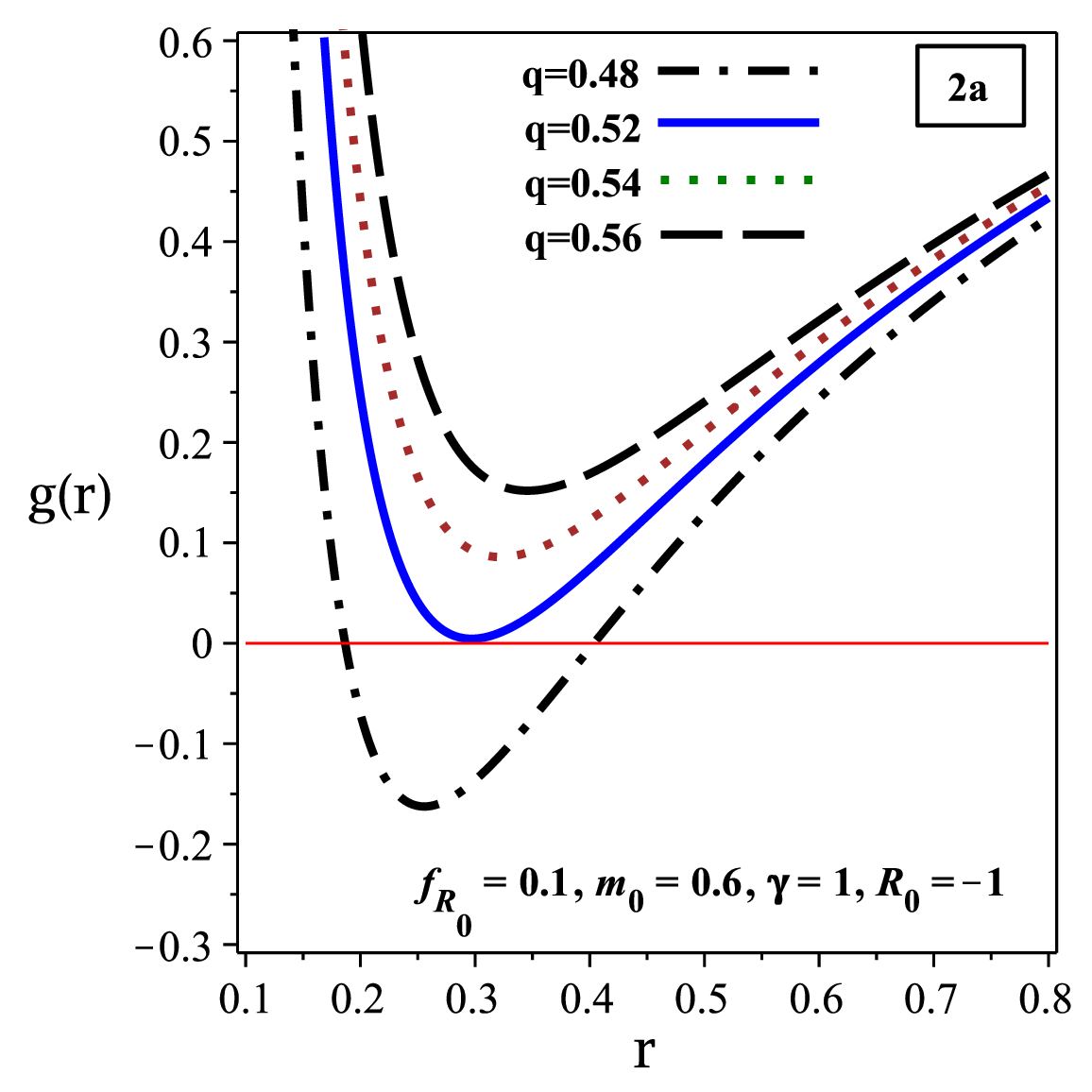} \includegraphics[width=0.35	%
\linewidth]{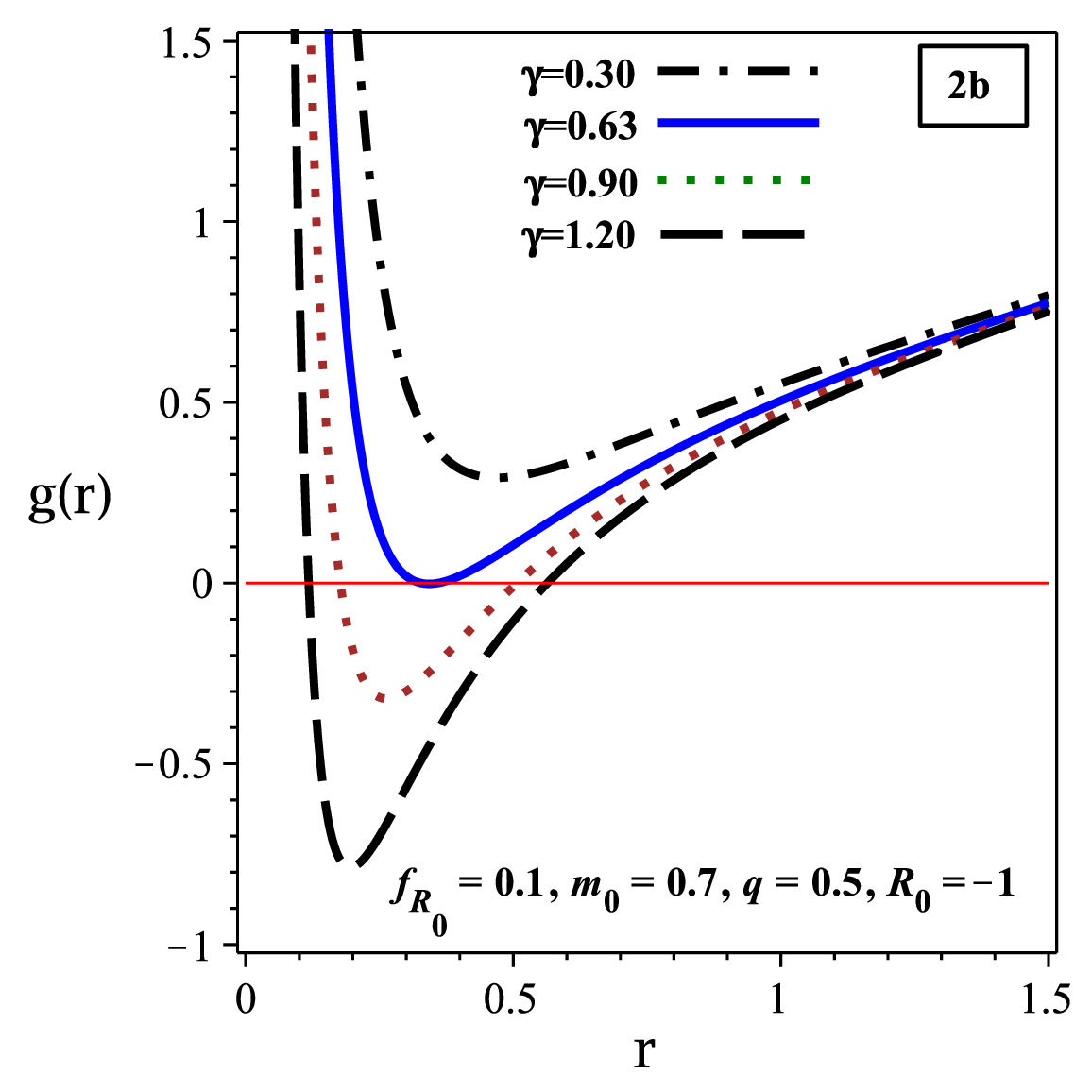}\newline
\includegraphics[width=0.35\linewidth]{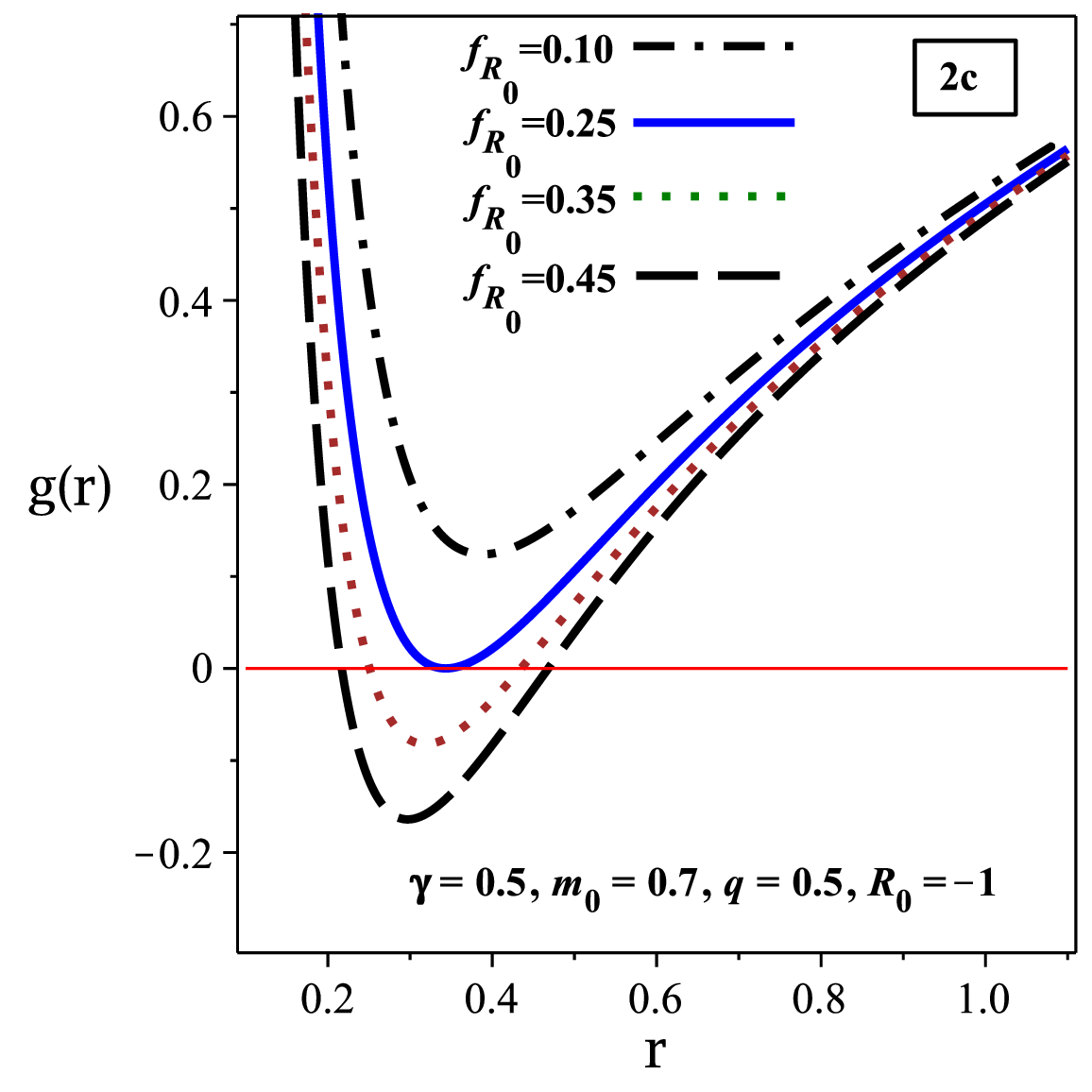} \includegraphics[width=0.35	%
\linewidth]{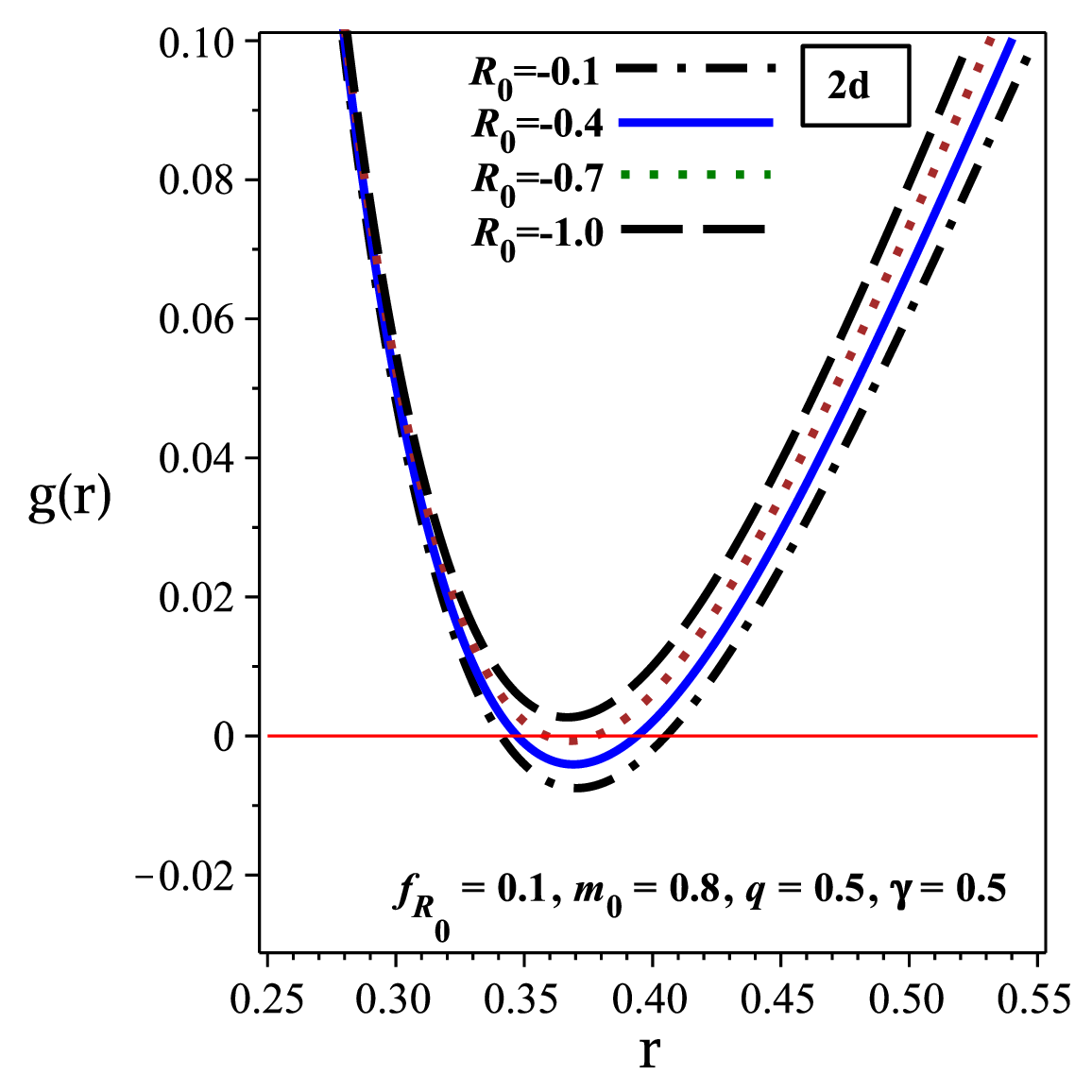}\newline
\caption{The function $g(r)$ versus $r$ is plotted for various parameter
values.}
\label{Fig2}
\end{figure}

We now have the opportunity to examine how various parameters in $F(R)-$%
ModMax theory influence the event horizon of black holes. Specifically, we
will evaluate the effects of electrical charge ($q$), the ModMax parameter ($%
\gamma$), and the parameters of $F(R)$ ($f_{R_{0}}$ and $R_{0}$) on this
type of black hole. Our findings are as follows:

i) The influence of electrical charge shows that increasing $q$ leads to a
decrease in the number of roots. This means that a black hole with a higher
charge in $F(R)-$ModMax theory does not possess an event horizon, resulting
in a naked singularity (see Fig. \ref{Fig2}a).

ii) The ModMax parameter demonstrates that increasing $\gamma$ results in
both a greater number of roots and a larger radius of the event horizon.
Specifically, a black hole with a larger $\gamma$ has two roots (see Fig. %
\ref{Fig2}b).

iii) Figure \ref{Fig2}c illustrates the impact of $f_{R_{0}}$ on the black
holes in $F(R)-$ModMax theory. It shows that as $f_{R_{0}}$ increases, both
the number of roots and the radius of the black holes also increase.

iv) The effect of $R_{0}$ is depicted in Fig. \ref{Fig2}d. The behavior of
black holes in relation to this parameter mirrors that of electrical charge;
specifically, as $\left| R_{0} \right|$ increases, the number of roots and
the radius of the black hole decrease.

Comparing the behavior of $q$ and $\gamma$ on the event horizon of these
black holes reveals that the ModMax parameter behaves oppositely to the
electric charge.

\section{Thermal stability}

In this section, we will derive the Hawking temperature and entropy for
accelerating ModMax AdS black holes within the framework of $F(R)$ gravity.
We will also examine local stability by analyzing the heat capacity.

\subsection{Temperature}

In the context of black holes, it was noted that a temperature of $T=0$
serves as a boundary between physical black holes ($T>0$) and non-physical
black holes ($T<0$) \cite{PanahPLB}. To identify the physical black holes,
we first calculate their Hawking temperature, and then we analyze the roots,
as well as the positive and negative regions of the Hawking temperature.

We calculate the Hawking temperature for accelerating ModMax black holes in $%
F(R)$ gravity by using the surface gravity. We use the following definition: 
\begin{equation}
T=\frac{\kappa }{2\pi }=\frac{\sqrt{\frac{-1}{2}\left( \nabla _{\mu }\chi
_{\nu }\right) \left( \nabla ^{\mu }\chi ^{\nu }\right) }}{2\pi },
\label{Temp}
\end{equation}%
where $\kappa $ is the surface gravity. In addition, $\chi =\partial _{t}$
is the Killing vector. Using the $C-$metric (\ref{Metric}), the surface
gravity is given 
\begin{equation}
\kappa =\sqrt{\frac{-1}{2}\left( \nabla _{\mu }\chi _{\nu }\right) \left(
\nabla ^{\mu }\chi ^{\nu }\right) }=\left. \frac{g^{\prime }\left( r\right) 
}{2}\right\vert _{r=r_{+}},  \label{k1}
\end{equation}%
where $r_{+}$ is the radius of the event horizon.

By considering Eqs. (\ref{Temp}) and (\ref{k1}), we find that the Hawking
temperature is 
\begin{equation}
T=\left. \frac{g^{\prime }\left( r\right) }{4\pi }\right\vert _{r=r_{+}}.
\label{TH2}
\end{equation}

To determine the Hawking temperature of accelerating ModMax black holes in $%
F(R)$ gravity (Eq. (\ref{g(r)F(R)})), we need to express the mass ($m_{0}$)
in terms of the scalar curvature ($R_{0}$), the event horizon radius ($r_{+}$%
), the ModMax parameter ($\gamma $), the $F(R)$ gravity term ($f_{R_{0}}$),
acceleration ($A$), and electric charge ($q$). To achieve this, we solve the
equation $g(r)=0$, which leads to 
\begin{equation}
m_{0}=\frac{\left( 1-\left( \frac{R_{0}}{12}+A^{2}\right) r_{+}^{2}\right)
r_{+}}{2\left( 1-A^{2}r_{+}^{2}\right) }+\frac{\left( 1-A^{2}\left(
1+f_{R_{0}}\right) r_{+}^{2}\right) q^{2}e^{-\gamma }}{2\left(
1+f_{R_{0}}\right) \left( 1-A^{2}r_{+}^{2}\right) r_{+}},  \label{mm}
\end{equation}%
by using the obtained metric function (\ref{g(r)F(R)}) and substituting the
mass (\ref{mm}) into equation (\ref{TH2}), one can derive the Hawking
temperature for these black holes in the following form: 
\begin{eqnarray}
T &=&\frac{\left( \frac{A^{2}r_{+}^{2}}{3}-1\right) R_{0}r_{+}}{16\pi \left(
1-A^{2}r_{+}^{2}\right) }-\frac{\left( 1+A^{4}\left( 1+f_{R_{0}}\right)
r_{+}^{4}-A^{2}\left( 2-f_{R_{0}}\right) r_{+}^{2}\right) q^{2}e^{-\gamma }}{%
4\pi \left( 1+f_{R_{0}}\right) \left( 1-A^{2}r_{+}^{2}\right) r_{+}^{3}} 
\notag \\[1pt]
&&  \notag \\
&&+\frac{\left( 1-A^{2}r_{+}^{2}\right) }{4\pi r_{+}},  \label{THfinal}
\end{eqnarray}%
where the Hawking temperature of these black holes (i.e., Eq. (\ref{THfinal}%
)) depends on several parameters, including the scalar curvature, the event
horizon, electric charge, and the parameters of ModMax, $F(R)$ gravity, and
acceleration.

Our analysis indicates that there are some interesting behaviors regarding
the temperature of accelerating ModMax AdS black holes in $F(R)$ gravity,
which are

i) There is a divergence point for the temperature at $r_{+} = \frac{1}{A}$,
which depends solely on the acceleration parameter. To avoid this
singularity, we do not allow $r_{+} = \frac{1}{A}$.

ii) The high-energy behavior of the temperature is given by 
\begin{equation}
\underset{r\rightarrow 0}{\lim }T\propto \frac{1}{4\pi r_{+}}\left( 1+\frac{%
q^{2}e^{-\gamma }\left( A^{2}\left( 1-f_{R_{0}}\right) r_{+}^{2}-1\right) }{%
\left( 1+f_{R_{0}}\right) r_{+}^{2}}\right) ,
\end{equation}%
where depends on the parameters $A$, $q$, $\gamma$, and $f_{R_{0}}$. Our
findings indicate that the Hawking temperature of small black holes can be
either positive ($T > 0$) or negative ($T < 0$), influenced by these various
parameters.

iii) The asymptotic behavior of temperature is determined as 
\begin{equation}
\underset{r\rightarrow \infty }{\lim }T\propto \frac{-\left(
R_{0}+12A^{2}\right) }{48\pi },
\end{equation}%
where indicates that very large black holes have a positive temperature when 
$A < \sqrt{\frac{-R_{0}}{12}}$. In other words, the temperature of the large
accelerating ModMax AdS black holes in $F(R)$ gravity is positive under the
same condition.

\begin{figure}[tbph]
\centering
\includegraphics[width=0.32\linewidth]{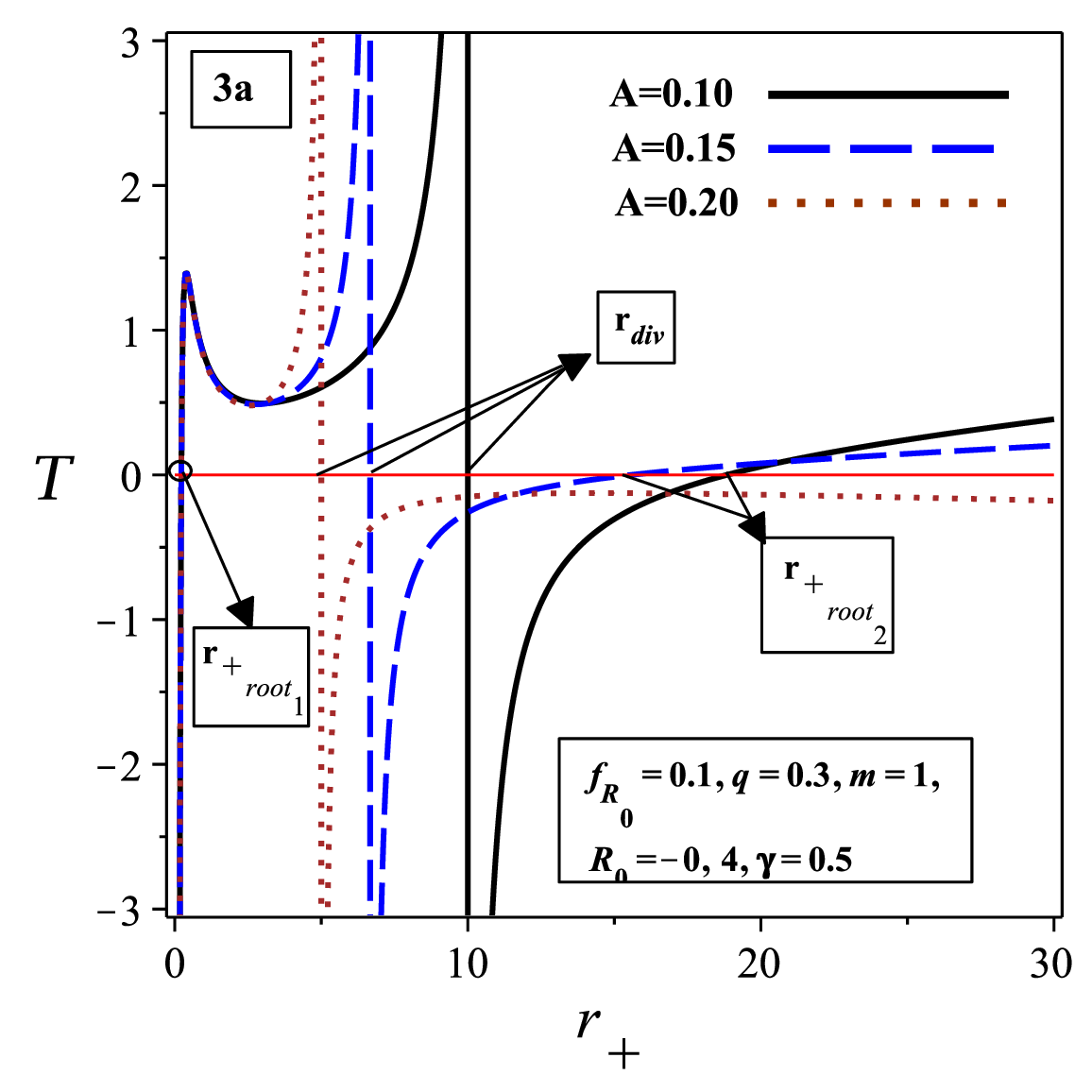} \includegraphics[width=0.32%
\linewidth]{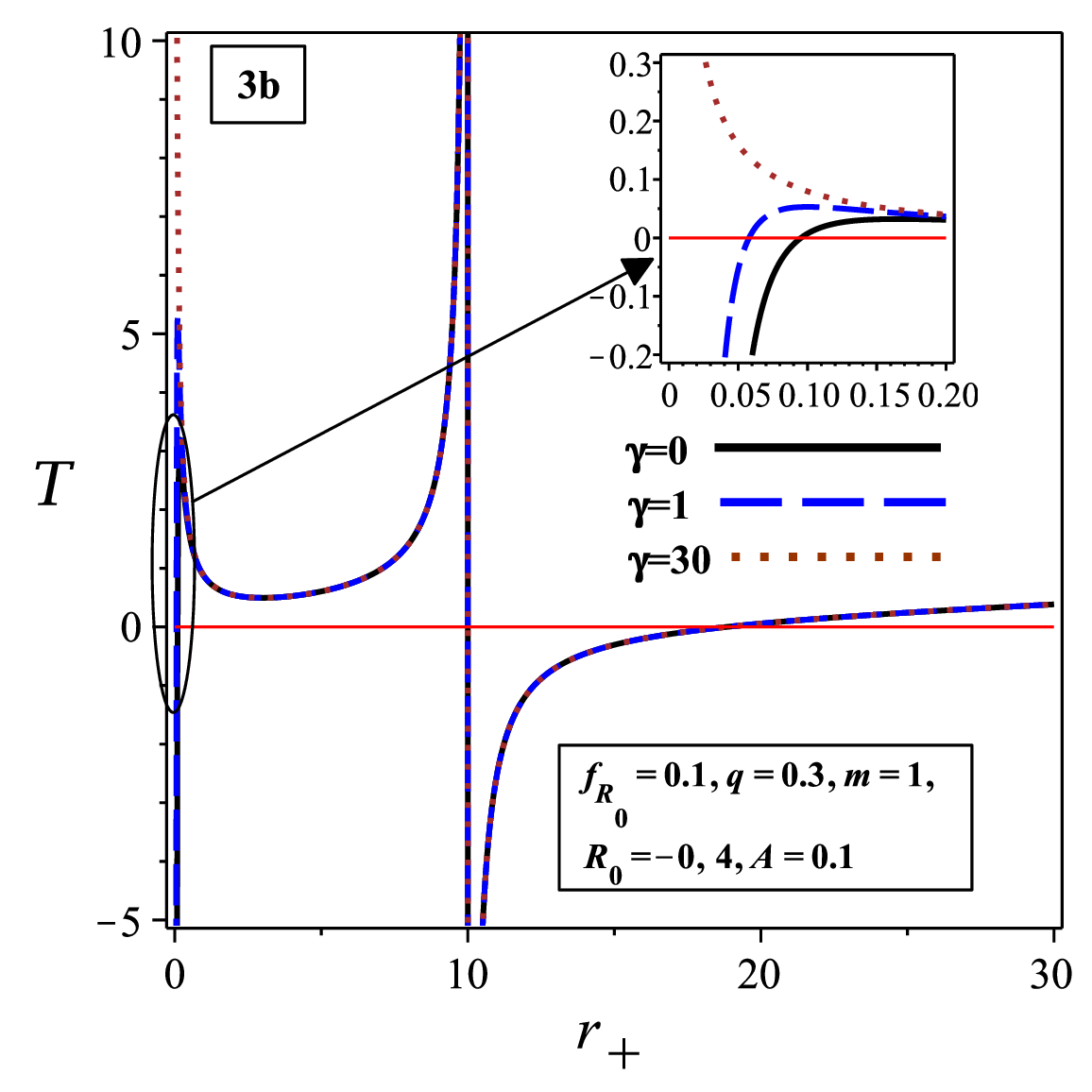} \includegraphics[width=0.32\linewidth]{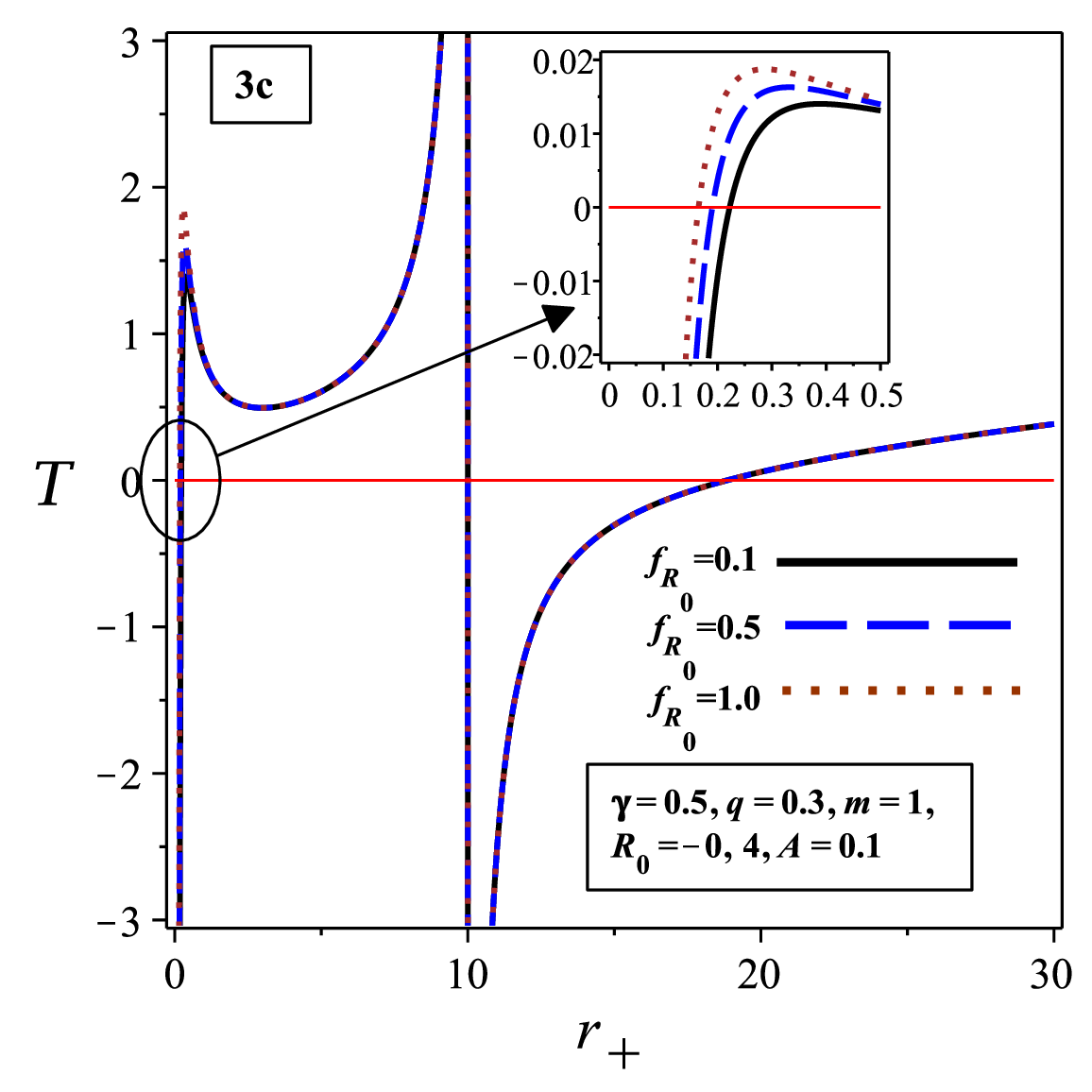} \newline
\caption{The Hawking temperature $T$ (Eq.(\protect\ref{THfinal})) versus $%
r_{+}$ is plotted for various parameter values. $T$ is positive in the
ranges $r_{+_{root_{1}}}<r_{+}<r_{+_{{div}}}$, and $r_{+}>r_{+_{root_{2}}}$
when $A<A_{crit}$. }
\label{Fig3}
\end{figure}

For more clarity, we present a plot of the Hawking temperature versus the
event horizon in Fig. \ref{Fig3}. Our results show that the temperature may
have two roots ($r_{+_{root_{1}}}$ and $r_{+_{root_{2}}}$) and one
divergence point ($r_{+_{{div}}}$). We analyze the temperature in greater
detail across three panels in Fig. \ref{Fig3}. The left panel of Fig. \ref%
{Fig3} (or Fig. \ref{Fig3}a) highlights two notable behaviors: i) a critical
value for the acceleration parameter ($A_{crit}=\sqrt{\frac{-R_{0}}{12}}$),
where large AdS black holes exhibit negative temperatures when $A > A_{crit}$%
. Conversely, for $A < A_{crit}$, the Hawking temperature of these black
holes remains positive. ii) The temperature has two roots and one divergence
point. Specifically, it is negative in the ranges $r_{+} < r_{+_{root_{1}}}$
and $r_{+_{{div}}} < r_{+} < r_{+_{root_{2}}}$. The temperature is positive
in the range $r_{+_{root_{1}}} < r_{+} < r_{+_{{div}}}$. Furthermore, when
considering $A < A_{crit}$, the temperature remains positive for $r_{+} >
r_{+_{root_{2}}}$. In summary, there are two regions where the temperature
can be positive when $A < A_{crit}$.

In Fig. \ref{Fig3}b, we identify another critical point related to the
ModMax parameter. There exists a critical value, denoted as $\gamma_{crit}$,
such that when $\gamma > \gamma_{crit}$, the Hawking temperature of small
black holes is positive. Conversely, for $\gamma < \gamma_{crit}$, these
small black holes cannot be considered physical systems. It is important to
note that the singularity of temperature is independent of the ModMax
parameter.

We examine how the parameter $f_{R_{0}}$ in $F(R)$ gravity affects the
Hawking temperature of black holes. To illustrate this, we plotted the
temperature against the event horizon for various values of $f_{R_{0}}$ in
Fig. \ref{Fig3}c. It is evident that the temperature's singularity is
independent of $f_{R_{0}}$. However, the smaller root of the temperature is
influenced by $f_{R_{0}}$. Specifically, as $f_{R_{0}}$ increases, the
physical area grows because the smaller root of the temperature decreases
with rising $f_{R_{0}}$.

\subsection{Entropy}

Another quantity that is useful for studying local stability is related to
entropy. Therefore, we need to calculate the entropy of accelerating ModMax
black holes in $F(R)$ gravity. However, we cannot apply the area law in its
standard form $S=\frac{\tilde{A}}{4}$, where $\tilde{A}$ is the horizon
area. To address this, Cognola, Elizalde, Nojiri, Odintsov, and Zerbini
introduced a modification of the area law in the following form \cite%
{Cognola2005} 
\begin{equation}
S=\frac{\tilde{A}\left( 1+f_{R_{0}}\right) }{4},  \label{SFR}
\end{equation}%
so, for extracting the entropy, we first need to determine the horizon area.
The horizon area is defined as 
\begin{equation}
\tilde{A}=\left. \int_{0}^{2\pi }\int_{0}^{\pi }\sqrt{g_{\theta \theta
}g_{\varphi \varphi }}\right\vert _{r=r_{+}},  \label{A}
\end{equation}%
and by using the $C-$metric in Eq. (\ref{Metric}), we can obtain 
\begin{eqnarray}
\tilde{A} &=&\int_{0}^{2\pi }\int_{0}^{\pi }\left. \frac{r^{2}\sin \theta }{%
\mathcal{K}^{2}\left( r,\theta \right) K}d\theta d\varphi \right\vert
_{r=r_{+}}=\left. \frac{4\pi r^{2}}{\left( 1-A^{2}r^{2}\right) K}\right\vert
_{r=r_{+}}  \notag \\
&&  \notag \\
&=&\frac{4\pi r_{+}^{2}}{\left( 1-A^{2}r_{+}^{2}\right) +\left( 1-\left( 
\frac{R_{0}}{12}+A^{2}\right) r_{+}^{2}\right) Ar_{+}+\frac{Aq^{2}e^{-\gamma
}}{\left( 1+f_{R_{0}}\right) r_{+}}-\frac{A^{3}q^{2}e^{-\gamma }r_{+}}{%
\left( 1+f_{R_{0}}\right) }},  \label{A1}
\end{eqnarray}%
where for obtaining the horizon area, we substitute the mass (Eq. (\ref{mm}%
)) into the expression $K=1+2m_{0}A$. Notably, when we set $A=0$, the
horizon area $\tilde{A}$ simplifies to $\tilde{A}=4\pi r_{+}^{2}$.

By substituting the horizon area (Eq. (\ref{A1})) into Eq. (\ref{SFR}), we
determine that the entropy of accelerating ModMax black holes in $F(R)$
gravity is expressed as 
\begin{equation}
S=\frac{\pi r_{+}^{2}\left( 1+f_{R_{0}}\right) }{\left(
1-A^{2}r_{+}^{2}\right) +\left( 1-\left( \frac{R_{0}}{12}+A^{2}\right)
r_{+}^{2}\right) Ar_{+}+\frac{Aq^{2}e^{-\gamma }}{\left( 1+f_{R_{0}}\right)
r_{+}}-\frac{A^{3}q^{2}e^{-\gamma }r_{+}}{\left( 1+f_{R_{0}}\right) }},
\label{S}
\end{equation}%
where depends on all the parameters of $F(R)$ gravity when coupled with
ModMax NLED theory in the $C-$metric. Notably, in the absence of $A$, the
entropy (as shown in equation \ref{S}) simplifies to $S=\pi r_{+}^{2}\left(
1+f_{R_{0}}\right)$. Moreover, when both $A$ and $f_{R_{0}}$ are absent
(i.e., $A=f_{R_{0}}=0$), the entropy further reduces to $S=\pi r_{+}^{2}$,
as expected.

To find the high-energy behavior of the entropy, we series the entropy (\ref%
{S}) versus the event horizon ($r_{+}$) in the limit $r_{+}\rightarrow 0$.
Our analysis indicate that%
\begin{equation}
\underset{r_{+}\rightarrow 0}{\lim }S\rightarrow 0,
\end{equation}%
the high-energy limit of the entropy is zero.

We can get the asymptotic behavior of the entropy (\ref{S}) by considering
the limit $r_{+}\rightarrow 0$ for Eq. (\ref{S}), which leads to%
\begin{equation}
\underset{r_{+}\rightarrow \infty }{\lim }S\propto \frac{-\pi \left(
1+f_{R_{0}}\right) }{\left( \frac{R_{0}}{12}+A^{2}\right) A},
\end{equation}%
whereas $\left\vert f_{R_{0}}\right\vert <1$, so the asymptotic behavior of
the entropy can be positive provided $A<\sqrt{\frac{-R_{0}}{12}}$. This
imposes that we must consider the negative value for $R_{0}$.

\begin{figure}[tbph]
\centering
\includegraphics[width=0.35\linewidth]{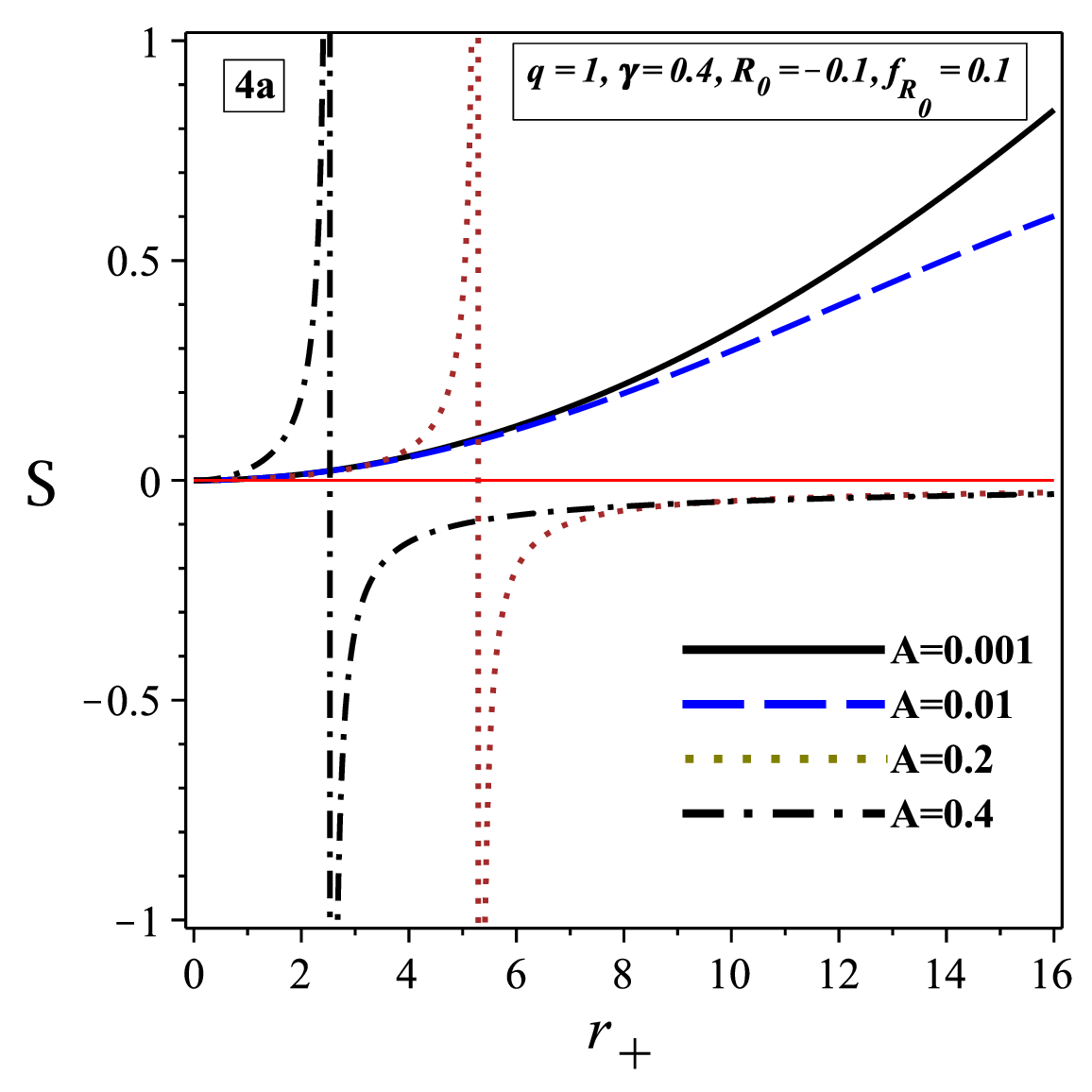} \includegraphics[width=0.35%
\linewidth]{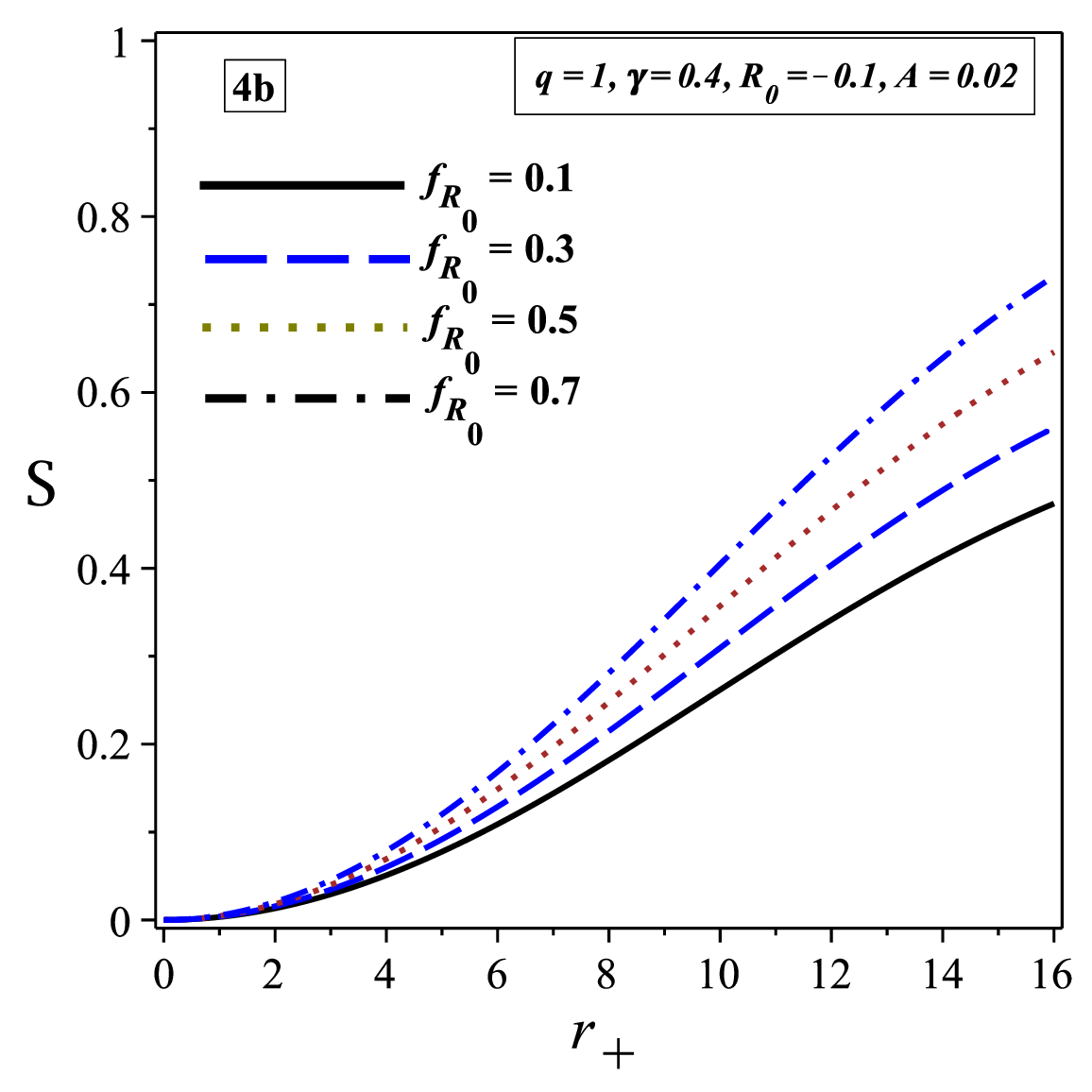} \newline
\caption{Entropy $S$ (Eq. (\protect\ref{S})) versus $r_{+}$ is plotted for
various parameter values.}
\label{Fig4}
\end{figure}

To study the effects of various parameters on the entropy of accelerating
ModMax black holes in $F(R)$ gravity, we plotted it versus the event horizon
in Fig. \ref{Fig4}.

Our findings in Fig. \ref{Fig4}a indicates that a divergence point ($r_{+_{{%
div}}}$) appears by increasing the parameter of accelerating. In addition,
this divergence point move to the small radius when $A$ increases. The
entropy is negative after this divergence point. Indeed, there is a
divergence point which constraint the parameters of accelerating black holes
in $F(R)-$ModMax theory. However, for small values of $A$, there is no
divergence point and the entropy is positive everywhere. In other words, the
accelerating ModMax black holes in $F(R)$ gravity can cover large black
holes when $A$ includes the small value. However the radius of accelearting
ModMax black hole reduces when $A$ includes the large value.

We plotted $S$ versus $r_{+}$ in Fig. \ref{Fig4}b, for different value of $%
f_{R_{0}}$. As one can see, the entropy increases by increasing $f_{R_{0}}$
and it is positive with any radius.

Our analysis from the entropy provide two constraints on the parameters of
accelerating and the scalar curvature as:

1) The large accelerating black holes in $F(R)-$ModMax theory include the
positive entropy when $A<\sqrt{\frac{-R_{0}}{12}}$. This imposes that $R_{0}$
must be negative.

2) To have the positive entropy (or avoid of the divergence point in the
entropy) of the accelerating ModMax black holes with arbitrary radius, we
must consider the small value for $A$.

\section{Heat capacity}

The heat capacity plays a significant role in determining thermal stability
of black holes in the context of the canonical ensemble. Indeed, we can
evaluate the local stability of black holes by studying the heat capacity to
determine thermal stability. Here, we discuss the thermal stability of
accelerating ModMax black holes in $F(R)$ gravity by using the heat capacity.

It is notable that, the study of heat capacity provides two fantastic
properties of black holes. These properties are:

i) \textbf{Phase transition points}: the divergences of the heat capacity
deremine the phase transition points (which is known as the second-order
phase transition points).

ii) \textbf{Thermal instability/stability}: in the canonical ensemble, the
positivity and negative of heat capacity reveal that the black hole is in a
thermally stable and unstable states, respectively.

For this purpose, we have to extract two important ponits of the heat
capacity. The first one is related to bound points, where the heat capacity
is zero. In bound point, the sign of temperature is changed. The sencond one
is related to phase transition ponits, where the heat capacity diverges.

The heat capacity with fixed charge ($C_{q}$) is defined in the following
form%
\begin{equation}
C_{q}=\frac{T}{\left( \frac{\partial T}{\partial S}\right) _{q}}.  \label{C}
\end{equation}

Now, we are able to extract the heat capacity with fixed charge by
considering the Hawking temperature (Eq. (\ref{THfinal})), and entropy (Eq. (%
\ref{S})) within Eq. (\ref{C}), which leads to

\begin{equation}
C_{q}=\frac{\left( 1-A^{2}r_{+}^{2}\right) \left( \frac{q^{2}e^{-\gamma }}{%
\left( 1+f_{R_{0}}\right) r_{+}^{2}}B_{4}-B_{5}\right) \left( B_{6}-\frac{%
Aq^{2}e^{-\gamma }\left( A^{2}r_{+}^{2}-3\right) }{\left( 1+f_{R_{0}}\right)
r_{+}}\right) \left( 1+f_{R_{0}}\right) \pi r_{+}^{2}}{\left( \frac{%
q^{2}e^{-\gamma }}{\left( 1+f_{R_{0}}\right) r_{+}^{2}}B_{7}+B_{8}\right)
\left( B_{9}-\frac{Aq^{2}e^{-\gamma }\left( 1-A^{2}r_{+}^{2}\right) }{\left(
1+f_{R_{0}}\right) r_{+}}\right) ^{2}},  \label{Cfinal}
\end{equation}%
where $B_{4}$, $B_{5}$, $B_{6}$, and $B_{7}$ are%
\begin{eqnarray}
B_{4} &=&\left( 1+f_{R_{0}}\right) A^{4}r_{+}^{4}-\left( 2-f_{R_{0}}\right)
A^{2}r_{+}^{2}+1,  \notag \\
&&  \notag \\
B_{5} &=&\left( A^{2}+\frac{R_{0}}{12}\right) A^{2}r_{+}^{4}-2\left( A^{2}+%
\frac{R_{0}}{8}\right) r_{+}^{2}+1,  \notag \\
&&  \notag \\
B_{6} &=&\left( A^{2}+\frac{R_{0}}{12}\right) Ar_{+}^{3}+Ar_{+}+2,  \notag \\
&&  \notag \\
B_{7} &=&\left( 1+f_{R_{0}}\right) A^{6}r_{+}^{6}-4\left( \frac{5}{4}%
-f_{R_{0}}\right) A^{4}r_{+}^{4}+\left( 7-f_{R_{0}}\right) A^{2}r_{+}^{2}-3,
\notag \\
&&  \notag \\
B_{8} &=&\left( A^{2}+\frac{R_{0}}{12}\right)
A^{4}r_{+}^{6}-A^{4}r_{+}^{4}-\left( A^{2}-\frac{R_{0}}{4}\right)
r_{+}^{2}+1,  \notag \\
&&  \notag \\
B_{9} &=&\left( A^{2}+\frac{R_{0}}{12}\right)
Ar_{+}^{3}+A^{2}r_{+}^{2}-Ar_{+}-1.
\end{eqnarray}

\begin{figure}[tbph]
\centering
\includegraphics[width=0.35\linewidth]{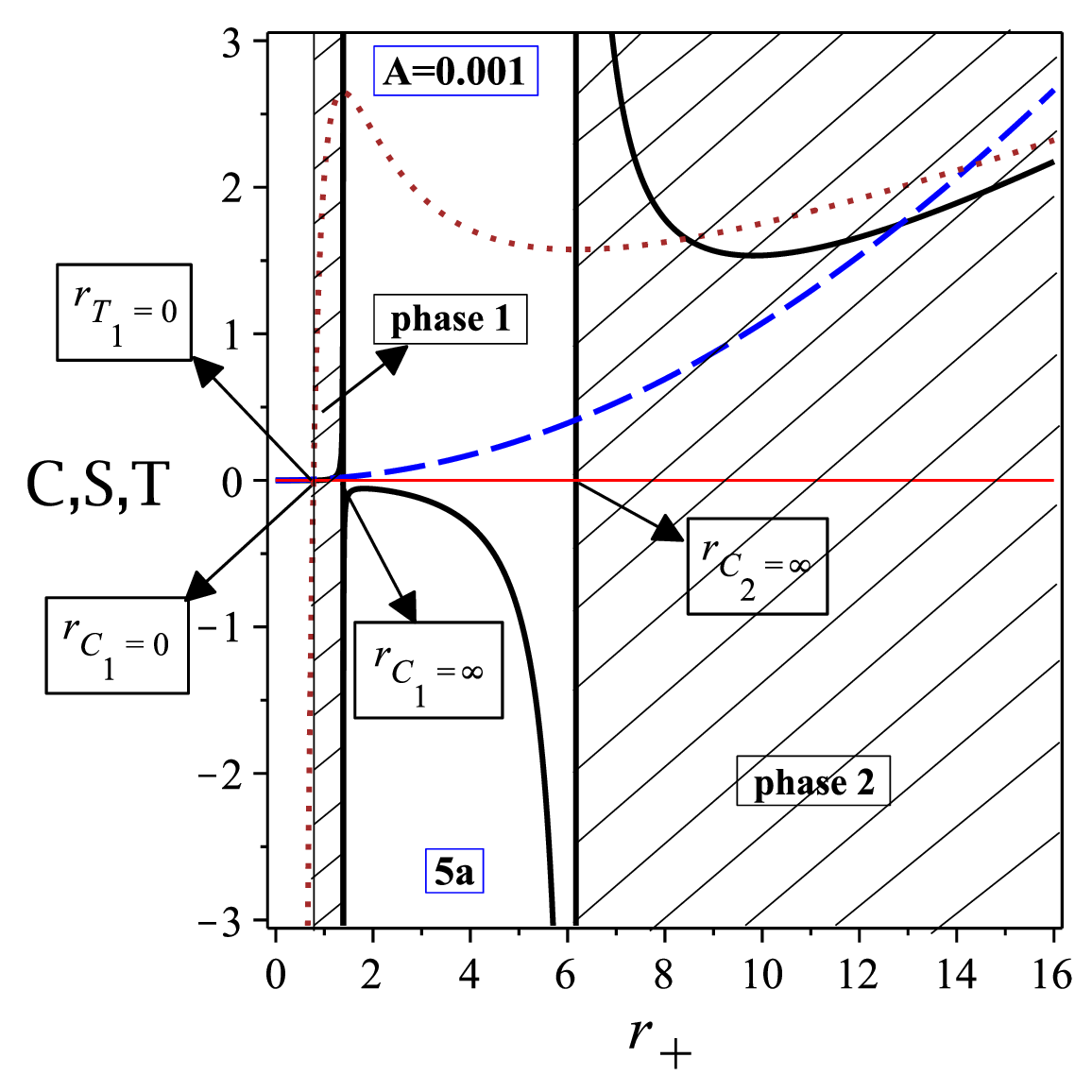} \includegraphics[width=0.35%
\linewidth]{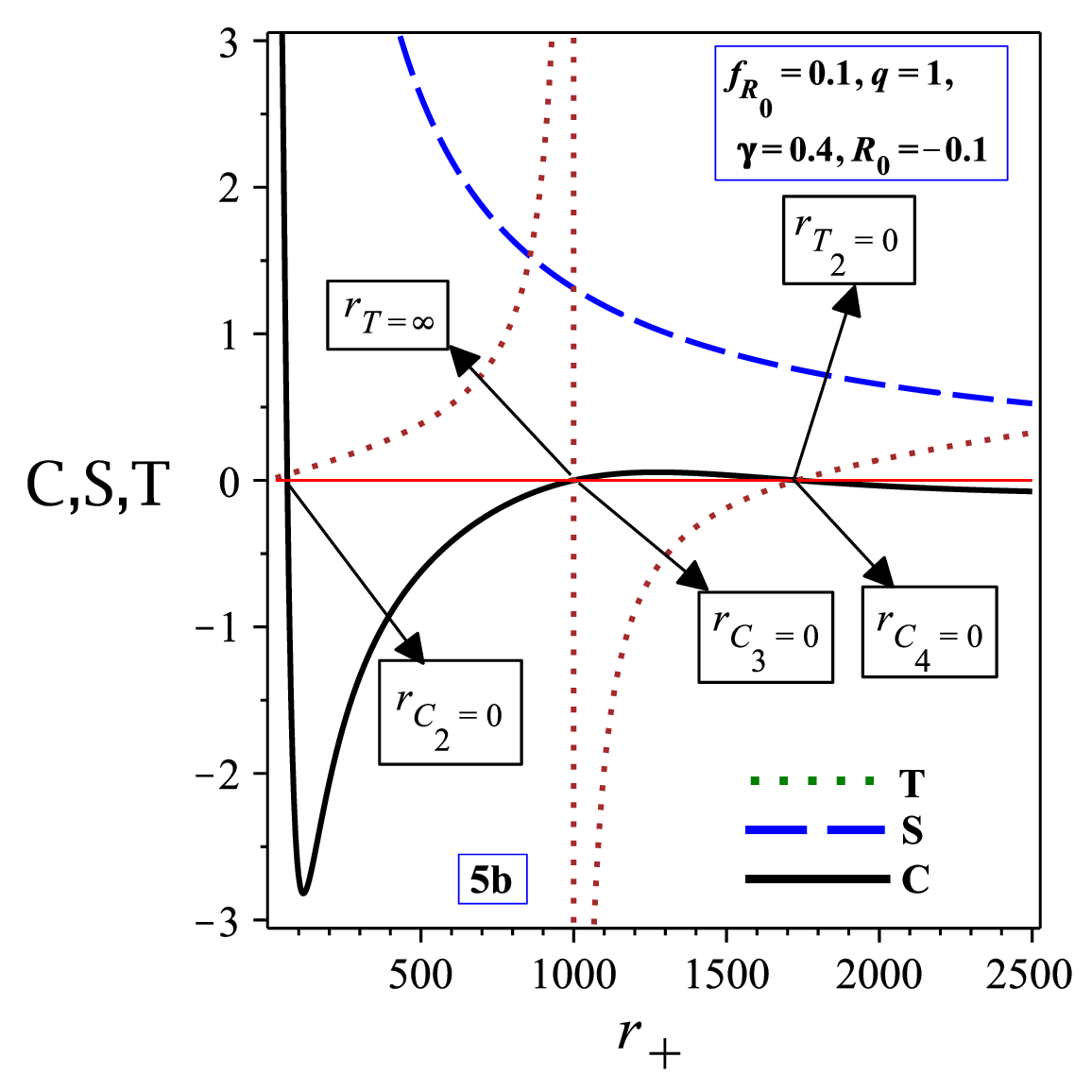} \newline
\caption{Entropy $S$, temperature $T$, and heat capacity $C$ versus $r_{+}$
is plotted for very small value of the accelerating parameter $A=0.001$. For
better visibility, we divide this figure into two panels with different
radii.}
\label{Fig5}
\end{figure}

\begin{figure}[tbph]
\centering
\includegraphics[width=0.35\linewidth]{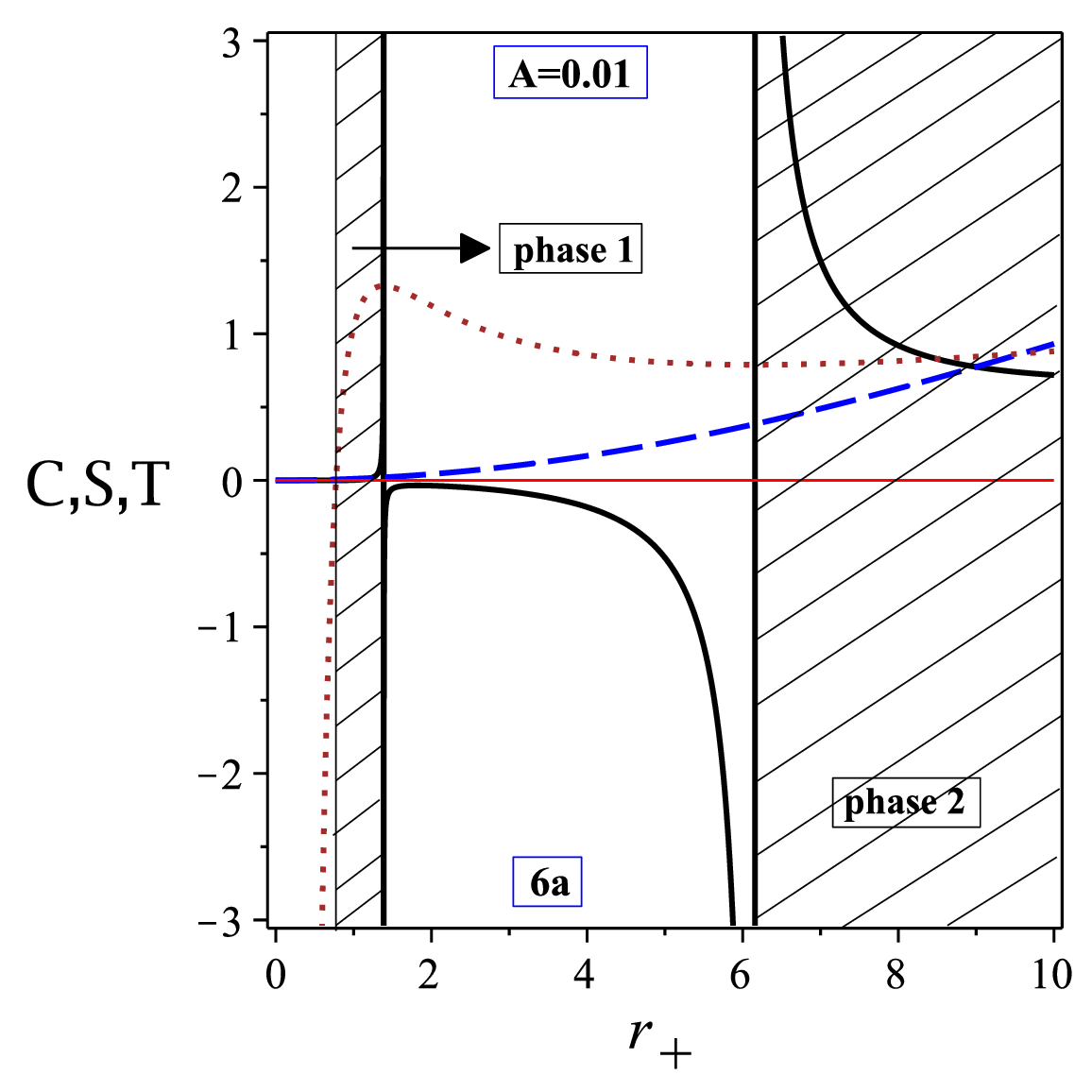} \includegraphics[width=0.35%
\linewidth]{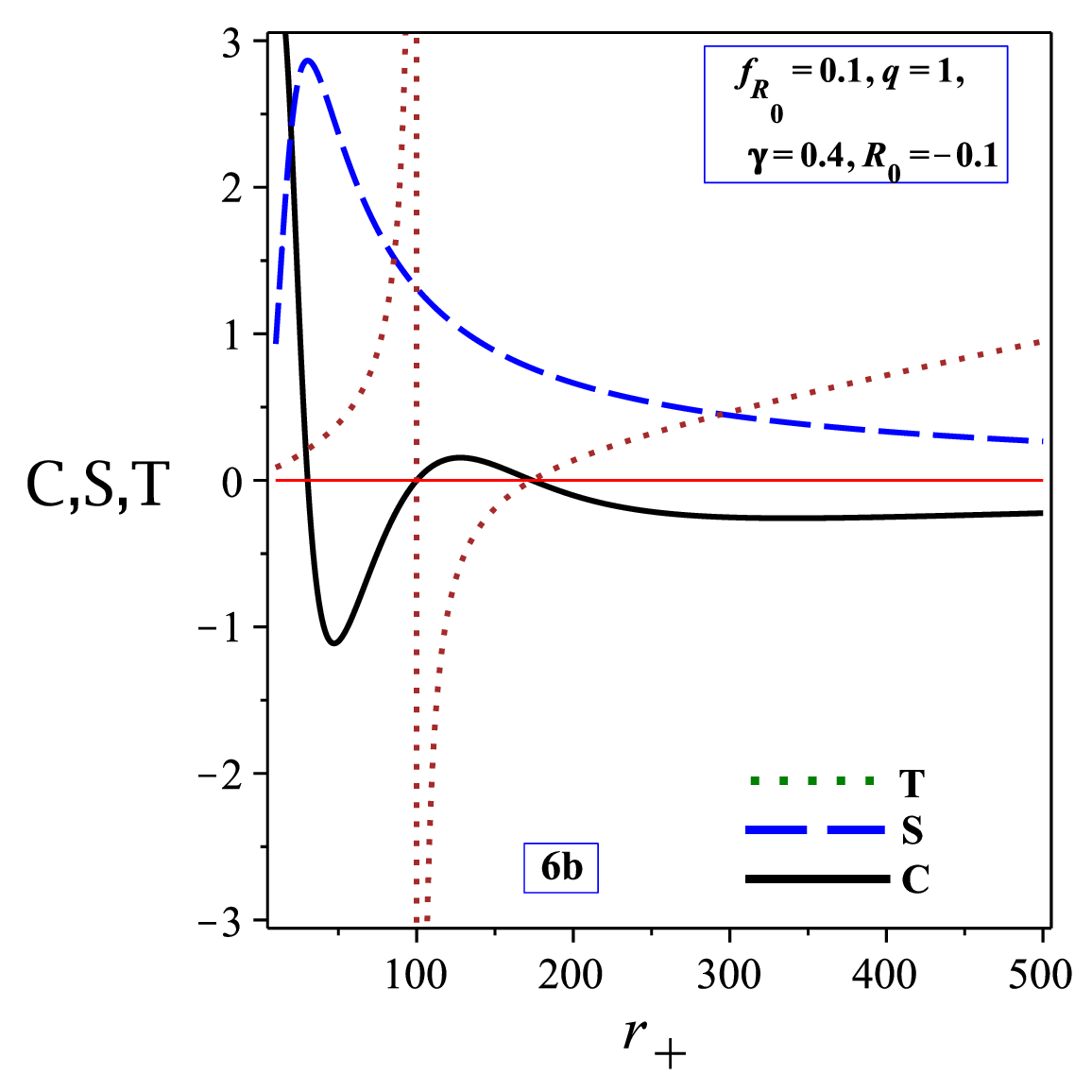} \newline
\caption{Entropy $S$, temperature $T$, and heat capacity $C$ versus $r_{+}$
is plotted for small value of the accelerating parameter $A=0.01$. For
better visibility, we divide this figure into two panels with different
radii.}
\label{Fig6}
\end{figure}

\begin{figure}[tbph]
\centering
\includegraphics[width=0.5\linewidth]{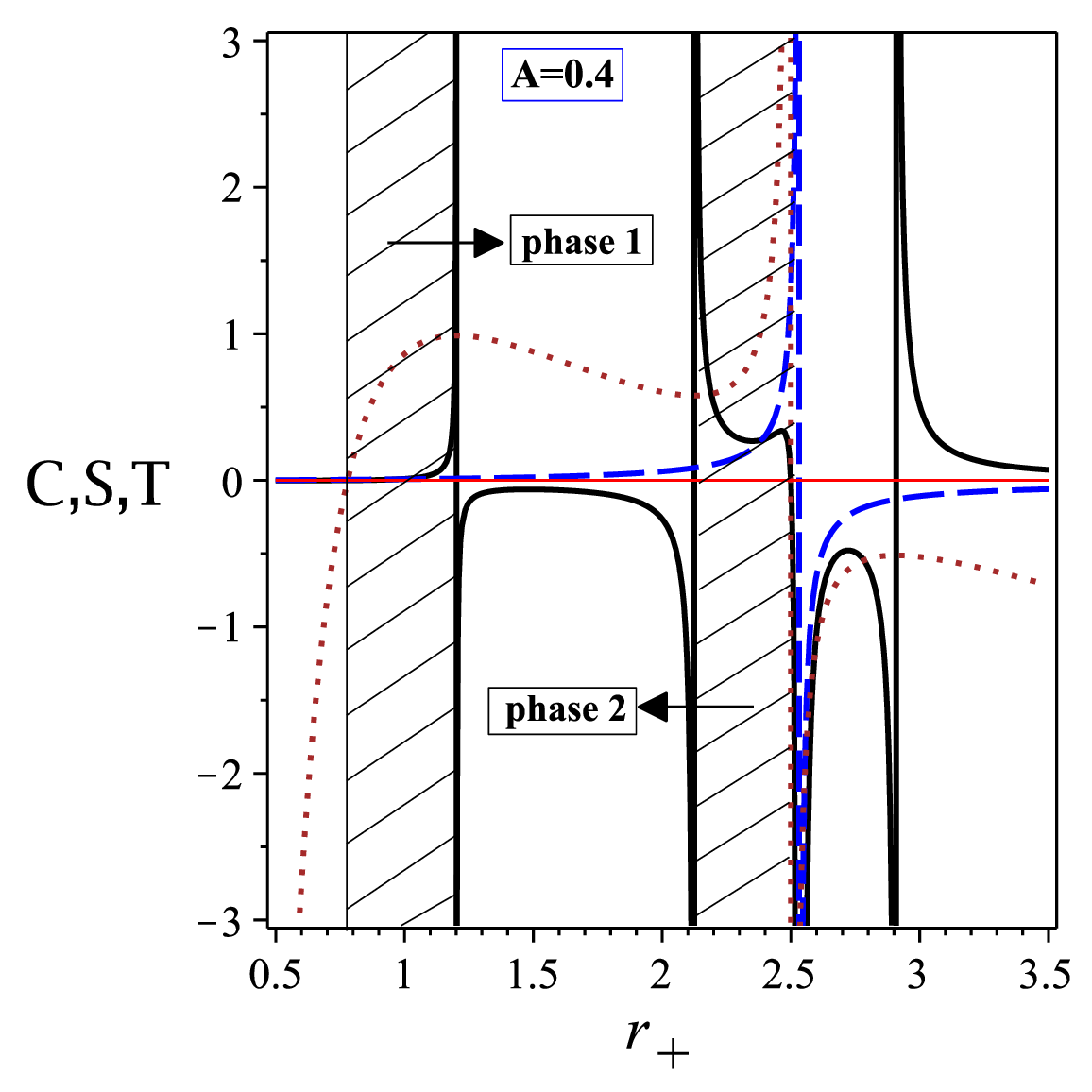} \newline
\caption{Entropy $S$, temperature $T$, and heat capacity $C$ versus $r_{+}$
is plotted for $A=0.4$.}
\label{Fig7}
\end{figure}

To determine the local stability we have to find the positive of the heat
capacity, temperature and entropy, simoultaneously. On the other hand, our
findings indicate that we should consider the small value for $A$, to avoid
of the divergences of the entropy and temperature. For this purpose, we
plotted the heat capacity, temperature and entropy versus the radius of
black holes in Figs. \ref{Fig5}-\ref{Fig7}.

Our fundings indicate that there are four roots for the heat capacity, which
are $r_{C_{1}=0}$, $r_{C_{2}=0}$, $r_{C_{3}=0}$, and $r_{C_{4}=0}$, from the
smallest to largest. In other words, $r_{C_{1}=0}$ devotes to the smallest
root of the heat capacity and $r_{C_{4}=0}$ is related to the largest root
of the heat capacity. In addition, there are two divergence points which
devote to $r_{C_{1}=\infty }$, and $r_{C_{2}=\infty }$. It is notable that $%
r_{C_{1}=\infty }$ and $r_{C_{2}=\infty }$ are related to the smallest and
largest divergence points, respectively.

In the previous subsection, we find two roots for the temperature which are $%
r_{T_{1}=0}$, and $r_{T_{2}=0}$. $r_{T_{1}=0}$ devotes to the smallest root
of the temperatutre and is the same with the smaleest root of the heat
capacity ($r_{C_{1}=0}$). Also $r_{T_{2}=0}$ is related to the largest root
of the temperature and is the same with the largest root of the heat
capacity ($r_{C_{4}=0}$). Where as we consider the small value for the
accelerating parameter, so the entropy is always positive.

Our analysis from the study the heat capacity, temperature and entropy of
accelerating ModMax black holes in $F(R)$ gravity reveal that seven
different regions, which are:

1) The first region devotes to the smallest accelearting ModMax black holes,
i.e, $r_{+}<r_{T_{1}=0}$. These black holes in this area are non-physical
objects.

2) The second region is located in $r_{T_{1}=0}<r_{+}<$ $r_{C_{1}=\infty }$.
Indeed, the accelearting ModMax black holes are physical objects and satisfy
the local condition (because the temperature, entropy and the heat capacity
are positive, simoultaneously) when their radius are in this area.

3) The third region belongs to $r_{C_{1}=\infty }<r_{+}<r_{C_{2}=\infty }$.
These black holes are unstable because the heat capacity is negative,
however their entropy and temperature ate positive.

4) The fourth region is located in $r_{C_{2}=\infty }<r_{+}<r_{C_{2}=0}$,
and the black holes are physical objects and satisfy the local stablity
condition because $T$, $S$, and $C$ are positive, simoulteneously.

5)\ The fifth region is in the range $r_{C_{2}=0}<r_{+}<r_{T=\infty }$. In
this area, the black holes are physical object but unstable.

6) The sixth region is determined between $r_{T=\infty }$ and the forth root
of the heat capacity (i.e., $r_{T=\infty }<r_{+}<r_{C_{4}=0}$). The
temperature of these black holes is negative and so we encounter with
non-physical objects.

7) The seventh region is $r_{+}>r_{C_{4}=0}$. The black holes are physical
objects but cannot satisfy the local stability condition.

So, the accelerating ModMax black holes in $F(R)$ garvity can satisfy the
physical and local stability conditions, simoultenously, when their radius
are in the second and forth regions. In other words, we encounter with
physical and stable black holes, simoultenuosly, when their radius are in
the ranges $r_{T_{1}=0}<r_{+}<$ $r_{C_{1}=\infty }$ (see the \textit{phase 1}
in Figs. \ref{Fig5}a and \ref{Fig6}a), and $r_{C_{2}=\infty
}<r_{+}<r_{C_{2}=0}$ (see the \textit{phase 2} in Figs. \ref{Fig5}a and \ref%
{Fig6}a).

Our findings about the effect of the accelearting parameter in Figs. \ref%
{Fig5}-\ref{Fig7} reveal that:

i) Increasing $A$ leads to decreases the physical and stable areas (the
second and the fourth regions).

ii) There is a phase transition between the second region (\textit{phase 1})
and the fourth region (\textit{phase 2}). In addition, this phase transition
removes when $A$ includes very large value.

\begin{figure}[tbph]
\centering
\includegraphics[width=0.35\linewidth]{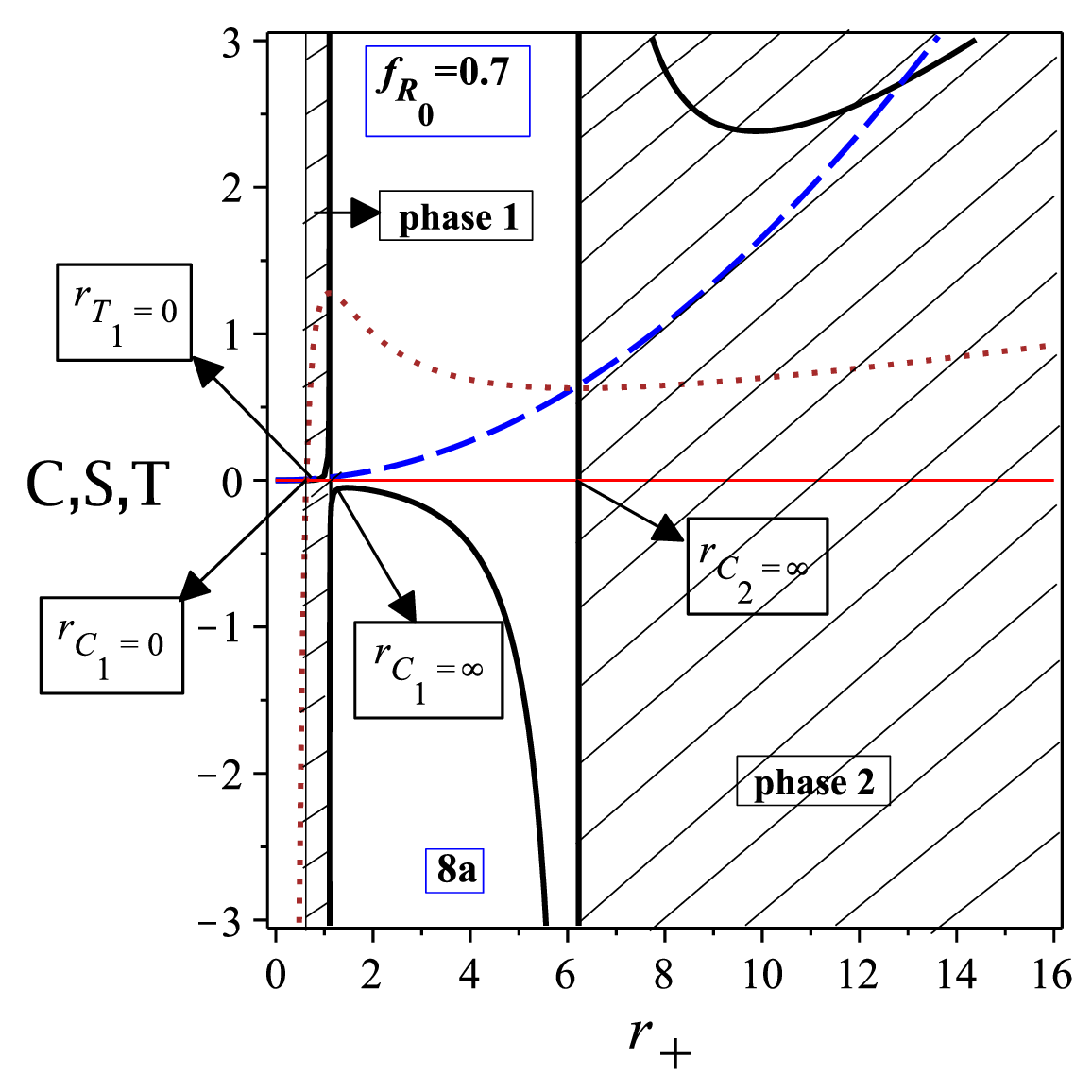} \includegraphics[width=0.35%
\linewidth]{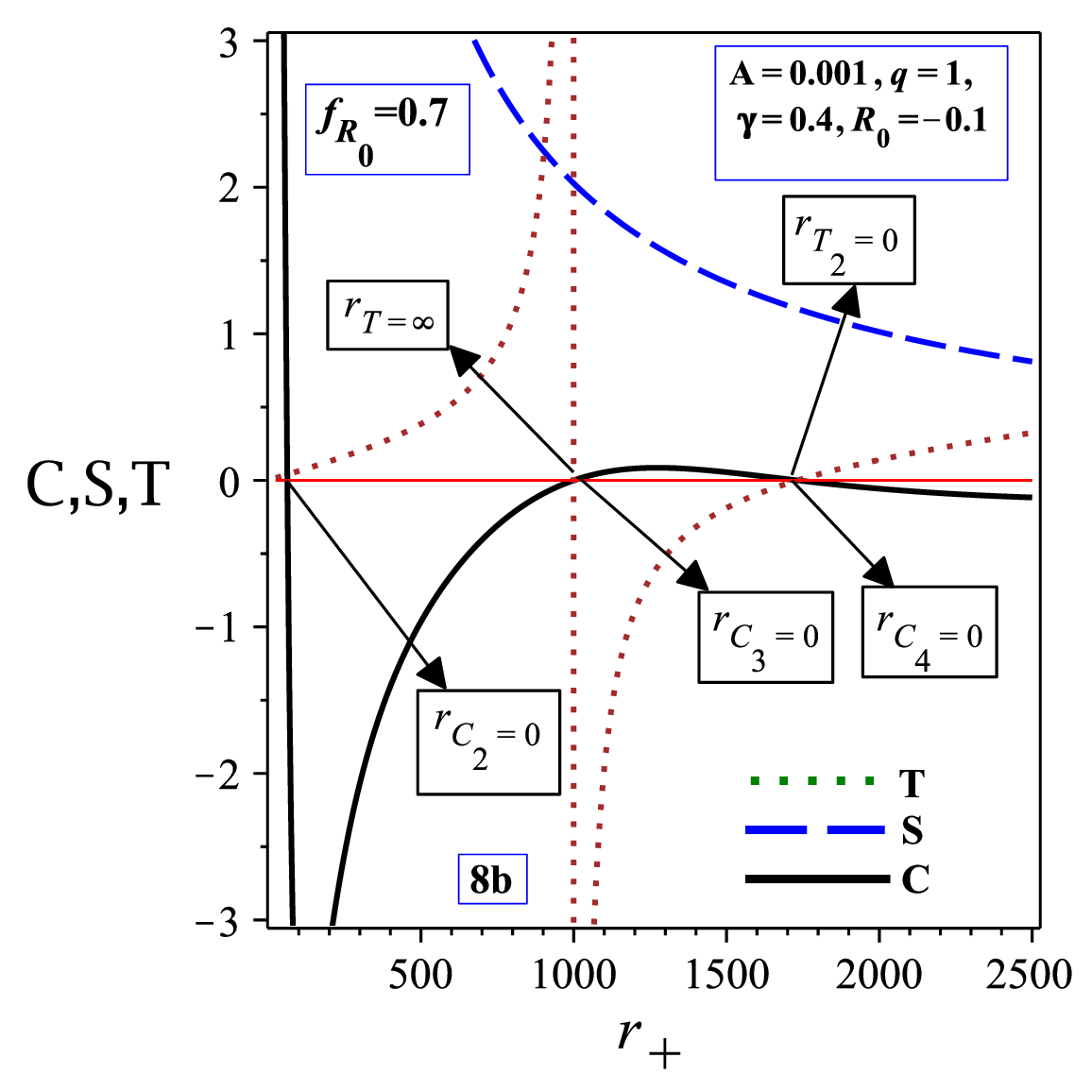} \newline
\caption{Entropy $S$, temperature $T$, and heat capacity $C$ versus $r_{+}$
is plotted for $f_{R_{0}}=0.7$. For better visibility, we divide this figure
into two panels with different radii.}
\label{Fig8}
\end{figure}

Now, we study the effect of the parameter of $F(R)$ gravity on these
quantities. By comparing between Fig. \ref{Fig5} and Fig. \ref{Fig8}, we
find that the second region (or $r_{T_{1}=0}<r_{+}<$ $r_{C_{1}=\infty }$)
decreases by increasing $f_{R_{0}}$.

\section{Optical appearance}


To study the optical signatures of the accelerating ModMax black hole in $%
F(R)$ gravity, we consider the dynamics of massless particles propagating
along null geodesics. The trajectories of photons encode essential
information about the causal structure of the spacetime and form the basis
for computing shadow observables. The Lagrangian describing particle motion
takes the form 
\begin{equation}
\mathcal{L} = \tfrac{1}{2} g^{\mu \nu} \dot{x}_{\mu} \dot{x}_{\nu}=\epsilon,
\end{equation}
In this framework, $\epsilon$ is equal to $0$ or $-1$ for massless or
massive particles, respectively, and $\dot{x}_{\mu}$ represents the photon
four-velocity with respect to the arbitrary affine parameter $\tau$.
Moreover, there exist conserved quantities associated with invariance under
time translations and rotations around the symmetry axis. These quantities
are identified with the photon energy $E$ and angular momentum $L$, which
take the form \cite{Grenzebach:2015oea,Zhang:2020xub,Karshiboev:2024xxx}

\begin{align}
E& =-g_{t\mu }\dot{x}^{\mu }=\frac{g(r)}{\mathcal{K}(r,\theta )^{2}}\dot{t},
\\
&  \notag \\
L& =g_{\varphi \mu }\dot{x}^{\mu }=\frac{X(\theta )\sin ^{2}\theta }{%
\mathcal{K}(r,\theta )^{2}K^{2}}\dot{\varphi}.
\end{align}

The function $g(r)$ characterizes the radial sector of the geometry, whereas 
$X(\theta)$ and $\mathcal{K}(r,\theta)$ encode angular contributions
influenced by acceleration effects and NLED.

For computational purposes, it is convenient to employ the \textit{Mino}
parameter $\sigma$, which regularizes the geodesic motion. Its relation to
the affine parameter $\tau$ is given by \cite%
{Frost:2020zcy,Frost:2021bvt,Batic:2021tjh}

\begin{equation}
\frac{d\sigma}{d\tau} = \frac{\mathcal{K}(r,\theta)^{2}}{r^{2}}.
\end{equation}

This change of parametrization proves particularly effective for analyzing
photon dynamics in accelerating backgrounds, since it allows a clean
separation between the radial and angular equations. 
In addition to $E$ and $L$, the separability of the Hamilton-Jacobi equation
introduces Carter's constant $Q$ \cite{Carter:1968rr}, which captures hidden
symmetries of the background. For null geodesics with $\epsilon =0$, the
Hamilton-Jacobi formalism \cite{Lim:2020bdj} yields the following set of
first-order differential equations 
\begin{align}
\frac{dt}{d\sigma }& =\frac{r^{2}E}{g(r)},  \label{eq:rmotion} \\
&  \notag \\
\frac{d\varphi }{d\sigma }& =\frac{LK^{2}}{X(\theta )\sin ^{2}\theta }, \\
&  \notag \\
\left( \frac{d\theta }{d\sigma }\right) ^{2}& =X(\theta )Q-\frac{L^{2}K^{2}}{%
\sin ^{2}\theta }\equiv \mathcal{T}(\theta ),  \label{eq:thetamotion} \\
&  \notag \\
\left( \frac{dr}{d\sigma }\right) ^{2}& =r^{4}E^{2}-r^{2}g(r)Q\equiv 
\mathcal{R}(r).
\end{align}

The functions $\mathcal{T}(\theta)$ and $\mathcal{R}(r)$ fully determine the
angular and radial motion of photons and play a central role in identifying
the photon sphere. Unstable circular photon orbits are characterized by the
photon radius $r_{ph}$ and photonic cone angle $\theta_{ph}$. These orbits
are determined by the extremity conditions of the following functions \cite%
{Zhang:2020xub}

\begin{align}
\mathcal{R}(r_{ph})& =0,~~~\&~~~\mathcal{R}^{\prime }(r_{ph})=0, \\
&  \notag \\
\mathcal{T}(\theta _{ph})& =0,~~~\&~~~\dot{\mathcal{T}}(\theta _{ph})=0.
\end{align}

Satisfying these relations ensures that photons are trapped on unstable
circular trajectories, which form the boundary of the black hole shadow.

It is convenient to express the results in terms of two dimensionless impact
parameters

\begin{equation}
\delta =\frac{\sqrt{Q}}{E},~~~\&~~~\zeta =\frac{L}{\sqrt{Q}},
\end{equation}

Solving the orbit conditions leads to compact expressions for the critical
parameters that determine the photon sphere

\begin{align}
\delta & =\frac{r_{ph}}{\sqrt{g(r_{ph})}},  \label{photonic1} \\
&  \notag \\
\zeta & =\frac{\sqrt{X(\theta _{ph})}\,\sin (\theta _{ph})}{K}.
\label{photonic2}
\end{align}

Eqs. \eqref{photonic1} and \eqref{photonic2} encapsulate the influence of
acceleration, ModMax NLED, and $F(R)$ curvature corrections on photon
dynamics. These quantities determine the photon sphere radius and thus set
the boundary of the observable black hole shadow.


\subsection{Photon Sphere Radius}

We can initiate the calculation of the photon sphere radius by rewriting the
radial equation of motion, given in Eq.~\eqref{eq:rmotion}, in the following
form

\begin{equation}
\frac{1}{Q r^4}\left( \frac{dr}{d\sigma}\right)^{2} + V_{\text{eff}}^r = 
\frac{E^{2}}{Q},
\end{equation}

where the effective potential $V_{\text{eff}}^r$ is defined as

\begin{equation}
V_{\text{eff}}^{r}=\left( \frac{1}{r^{2}}-A^{2}\right) \left( 1-\frac{2m_{0}%
}{r}+\frac{q^{2}e^{-\gamma }}{(1+f_{R_{0}})r^{2}}\right) -\frac{R_{0}}{12}.
\end{equation}

The radius of the photon sphere can be determined by imposing the conditions 
$\frac{dr}{d\sigma} = 0$ and $\frac{d^2 r}{d\sigma^2} = 0$, which are
equivalent to requiring that the effective potential $V_{\text{eff}}^r$
reaches an extremum. Therefore, the photon sphere radius $r_{ph}$ is
obtained by differentiating $V_{\text{eff}}^r$ with respect to $r$ and
setting the result equal to zero. This procedure leads to the following
cubic equation \cite{Frost:2020zcy,Frost:2021bvt,Heidari:2025llu}

\begin{equation}
r_{ph}^{3}-\frac{\left( A^{2}q^{2}e^{-\gamma }-(1+f_{R_{0}})\right) }{%
(1+f_{R_{0}})m_{0}A^{2}}r_{ph}^{2}-\frac{3}{A^{2}}r_{ph}+\frac{%
2q^{2}e^{-\gamma }}{(1+f_{R_{0}})m_{0}A^{2}}=0.  \label{photon}
\end{equation}

It is worth emphasizing that the above equation does not depend explicitly
on the constant scalar curvature $R_0$. However, both the acceleration
parameter $A$, the ModMax parameter $\gamma$, and $F(R)$ gravity play
crucial roles in determining the photon sphere radius. In order to explore
these effects in detail, we solve Eq.~\eqref{photon} numerically for
representative parameter choices. Specifically, we consider the case $m_0 =
1 $, $q = 0.3$, $R_0 = -0.4$, and investigate the dependence of $r_{ph}$ on
variations of the acceleration parameter $A$, the ModMax parameter $\gamma$,
and the $F(R)$ gravity parameter $f_{R_0}$. The corresponding numerical
results are summarized in Tables \ref{Tab:rph1} and \ref{Tab:rph2}. Table %
\ref{Tab:rph1} displays the behavior of the photon sphere radius for
different values of $A$ in the interval $0.0 \leq A \leq 0.9$, combined with
several representative values of the ModMax parameter $\gamma$, while
keeping $f_{R_0} = 0.1$ fixed. Table \ref{Tab:rph2} shows the variation of $%
r_{ph}$ for different values of $f_{R_0}$ in the range $0.0 \leq f_{R_0}
\leq 0.8$, together with different values of $A$, while fixing $\gamma = 0.1$%
. These results clearly illustrate the significant influence of the
acceleration parameter, the ModMax parameter, and the $F(R)$ gravity
parameter on the photon sphere radius.

\begin{table}[!ht]
\caption{Photon sphere radius $r_{ph}$ for $m_0 = 1$, $q = 0.3$, $R_0 = -0.4$%
, and $f_{R_0} = 0.1$ in the case of an accelerating ModMax black hole, for
different values of the acceleration parameter $A$ (ranging from $0.0$ to $%
0.9$) and the ModMax parameter $\protect\gamma$ ($0.0$, $0.5$, $1.0$, $1.5$,
and $10.0$).}
\label{Tab:rph1}\centering
\begin{tabular}{|c|c|c|c|c|c|}
\hline
A & ~~$\gamma$ = 0.0~~ & $~~\gamma$ = 0.5 ~~ & ~~$\gamma$ = 1.0~~ & $%
~~\gamma $ = 1.5 ~~ & ~~$\gamma$ = 10.0~~ \\ \hline\hline
0.00 & 2.9444 & 2.9665 & 2.9798 & 2.9878 & 3.0000 \\ \hline
0.01 & 2.9436 & 2.9657 & 2.9789 & 2.9869 & 2.9991 \\ \hline
0.05 & 2.9233 & 2.9450 & 2.9580 & 2.9658 & 2.9778 \\ \hline
0.09 & 2.8780 & 2.8989 & 2.9114 & 2.9189 & 2.9304 \\ \hline
0.1 & 2.8632 & 2.8838 & 2.8962 & 2.9036 & 2.9150 \\ \hline
0.5 & 1.9788 & 1.9873 & 1.9924 & 1.9954 & 2.0000 \\ \hline
0.9 & 1.3962 & 1.3993 & 1.4011 & 1.4022 & 1.4038 \\ \hline
\end{tabular}%
\end{table}

\begin{table}[!ht]
\caption{Photon sphere radius $r_{ph}$ for $m_0 = 1$, $q = 0.3$, $R_0 = -0.4$%
, and $\protect\gamma = 0.1$ for different values of the acceleration
parameter $A$ (ranging from $0.0$ to $0.9$) and the $F(R)$ gravity parameter 
$f_{R_0}$ ($0.0$, $0.2$, $0.4$, $0.6$, and $0.8$).}
\label{Tab:rph2}\centering
\begin{tabular}{|c|c|c|c|c|c|}
\hline
A & ~$f_{R_0}$ = 0.0~ & $~f_{R_0}$ = 0.2 ~ & ~$f_{R_0}$ = 0.4~ & $~f_{R_0}$
= 0.6 ~ & ~$f_{R_0}$ = 0.8~ \\ \hline\hline
0.00 & 2.9447 & 2.9541 & 2.9607 & 2.9657 & 2.9695 \\ \hline
0.01 & 2.9438 & 2.9532 & 2.9598 & 2.9648 & 2.9687 \\ \hline
0.05 & 2.9235 & 2.9327 & 2.9392 & 2.9441 & 2.9479 \\ \hline
0.09 & 2.8782 & 2.8871 & 2.8933 & 2.8980 & 2.9017 \\ \hline
0.1 & 2.8635 & 2.8722 & 2.8784 & 2.8830 & 2.8866 \\ \hline
0.5 & 1.9789 & 1.9825 & 1.9851 & 1.9870 & 1.9885 \\ \hline
0.9 & 1.3963 & 1.3976 & 1.3985 & 1.3992 & 1.3997 \\ \hline
\end{tabular}%
\end{table}


From Table~\ref{Tab:rph1}, it is evident that the photon sphere radius
decreases monotonically as the acceleration parameter $A$ increases,
regardless of the value of the ModMax parameter $\gamma$. For small values
of $A$, $r_{ph}$ is close to its Schwarzschild-like value ($r_{ph} \approx 3$
when $A = 0$ and $\gamma \to \infty$), while for large $A$ the radius is
significantly reduced. This indicates that acceleration tends to compress
the photon sphere toward the black hole horizon. Additionally, for fixed $A$%
, the parameter $\gamma$ produces a mild but noticeable effect: increasing $%
\gamma$ shifts the photon sphere outward, asymptotically approaching the
value $r_{ph} = 3$ in the large-$\gamma$ limit. This behavior highlights the
role of NLED in counteracting the reduction caused by acceleration.

Turning to Table~\ref{Tab:rph2}, we observe the influence of the $F(R)$
gravity parameter $f_{R_0}$. As with Table~\ref{Tab:rph1}, the acceleration
parameter $A$ continues to decrease the photon sphere radius. However, for
fixed $A$, increasing $f_{R_0}$ causes a small outward shift of the photon
sphere. This suggests that modifications of the underlying gravitational
theory through $F(R)$ corrections work in the opposite direction of
acceleration, expanding the photon sphere radius.

To investigate the behavior of the photonic cone in the context of
accelerating ModMax black holes within $F(R)$ gravity, we begin by rewriting
the $\theta $ motion equation from Eq.~\eqref{eq:thetamotion} in the
following form 
\begin{equation}
\frac{\sin ^{2}\theta }{Q}\left( \frac{d\theta }{d\sigma }\right)
^{2}+V_{\theta }=-\frac{L^{2}K^{2}}{Q},
\end{equation}%
where the corresponding effective potential is expressed as 
\begin{equation}
V_{\text{eff}}^{\theta }=-X(\theta )\sin ^{2}\theta =-\sin ^{2}\theta \left(
1+2m_{0}A\cos \theta +\frac{q^{2}e^{-\gamma }A^{2}\cos ^{2}\theta }{%
1+f_{R_{0}}}\right) .  \label{eq:Vtheta}
\end{equation}

The photonic cone arises when $\frac{d\theta }{d\sigma }=0$ and $\frac{%
d^{2}\theta }{d\sigma ^{2}}=0$, corresponding to the minimum of the
effective potential $V_{\text{eff}}^{\theta }$, and these conditions lead to
a cubic equation for the photonic cone $\theta _{ph}$ as follows 
\begin{equation}
x^{3}-\frac{3(1+f_{R_{0}})m_{0}}{2q^{2}e^{-\gamma }A}x^{2}-\frac{%
(1+f_{R_{0}})\left( 1+\frac{2q^{2}e^{-\gamma }A^{2}}{1+f_{R_{0}}}\right) }{%
2q^{2}e^{-\gamma }A^{2}}x+\frac{(1+f_{R_{0}})m_{0}}{2q^{2}e^{-\gamma }A}=0,
\end{equation}%
where $x=\cos {\theta }$ is treated as the variable. The numerical results
for the photonic cone angle $\theta _{ph}$ are displayed in Table~\ref%
{Tab:phtheta1} and Table~\ref{Tab:phtheta2}.

\begin{table}[!ht]
\caption{photonic cone angle $\protect\theta_{ph}$ for $m_0 = 1$, $q = 0.3$, 
$f_{R_0} = 0.1$, and $R_0 = - 0.4$ in the accelerating ModMax $F(R)$
gravity. The results are presented for different values of the acceleration
parameter $A$ (ranging from $0.01$ to $0.9$) and the ModMax parameter $%
\protect\gamma$ ($0.0$, $0.5$, $1.0$, $1.5$, and $10.0$).}
\label{Tab:phtheta1}\centering
\begin{tabular}{|c|c|c|c|c|c|}
\hline
A & ~~$\gamma$ = 0.0~~ & $~~\gamma$ = 0.5 ~~ & ~~$\gamma$ = 1.0~~ & $%
~~\gamma $ = 1.5 ~~ & ~~$\gamma$ = 10.0~~ \\ \hline\hline
0.01 & 0.496818 & 0.496818 & 0.496818 & 0.496818 & 0.496818 \\ \hline
0.05 & 0.484202 & 0.484199 & 0.484198 & 0.484197 & 0.484196 \\ \hline
0.09 & 0.472015 & 0.472002 & 0.471993 & 0.471988 & 0.471980 \\ \hline
0.1 & 0.469069 & 0.469051 & 0.469039 & 0.469032 & 0.469022 \\ \hline
0.5 & 0.393852 & 0.393060 & 0.392576 & 0.392282 & 0.391827 \\ \hline
0.9 & 0.366496 & 0.364596 & 0.363434 & 0.362727 & 0.361631 \\ \hline
\end{tabular}%
\end{table}

\begin{table}[!ht]
\caption{photonic cone angle $\protect\theta_{ph}$ for $m_0 = 1$, $q = 0.3$, 
$\protect\gamma = 0.1$, and $R_0 = - 0.4$ in the accelerating ModMax $F(R)$
gravity. The results are presented for different values of the acceleration
parameter $A$ (ranging from $0.01$ to $0.9$) and the $F(R)$ gravity
parameter $f_{R_0}$ ($0.0$, $0.2$, $0.4$, $0.6$, and $0.8$).}
\label{Tab:phtheta2}\centering
\begin{tabular}{|c|c|c|c|c|c|}
\hline
A & ~$f_{R_0}$ = 0.0~ & $~f_{R_0}$ = 0.2 ~ & ~$f_{R_0}$ = 0.4~ & $~f_{R_0}$
= 0.6 ~ & ~$f_{R_0}$ = 0.8~ \\ \hline\hline
0.01 & 0.496818 & 0.496818 & 0.496818 & 0.496818 & 0.496818 \\ \hline
0.05 & 0.484202 & 0.484201 & 0.484200 & 0.484200 & 0.484199 \\ \hline
0.09 & 0.472015 & 0.472009 & 0.472005 & 0.472002 & 0.472000 \\ \hline
0.1 & 0.469069 & 0.469061 & 0.469056 & 0.469051 & 0.469048 \\ \hline
0.5 & 0.393842 & 0.393509 & 0.393271 & 0.393091 & 0.392952 \\ \hline
0.9 & 0.366474 & 0.365675 & 0.365102 & 0.364671 & 0.364336 \\ \hline
\end{tabular}%
\end{table}

We note that the case of $A = 0$ is absent from Tables~\ref{Tab:phtheta1}
and~\ref{Tab:phtheta2}. According to Eq.~\eqref{eq:Vtheta}, when $A = 0$,
the photonic cone angle is fixed at $\theta_{ph} = \pi/2$, independent of
all other parameters. From Table~\ref{Tab:phtheta1}, it is observed that the
photonic cone angle decreases as the acceleration parameter $A$ increases.
For very small values of $A$, $\theta_{ph}$ remains nearly constant ($%
\theta_{ph} \approx 0.497$), almost independent of the ModMax parameter $%
\gamma$. However, as $A$ grows, the cone angle decreases significantly. The
role of $\gamma$ becomes more pronounced at larger $A$, with higher $\gamma$
values producing slightly smaller cone angles. This suggests that the NLED
corrections encoded in $\gamma$ enhance the effect of acceleration by
further narrowing the photonic cone.

Table~\ref{Tab:phtheta2} highlights the influence of the $F(R)$ gravity
parameter $f_{R_0}$. Similar to the case in Table~\ref{Tab:rph2}, the
acceleration parameter $A$ dominates the overall behavior, driving the
decrease of $\theta_{ph}$ as $A$ increases. However, for fixed $A$,
increasing $f_{R_0}$ produces a modest reduction in the photonic cone angle.

\subsection{Celestial Coordinates}

In order to determine the shadow cast by the black hole, we begin by
introducing an orthonormal tetrad basis associated with an observer located
at fixed coordinates $(r_{O},\theta_{O})$ \cite%
{Li:2020drn,Grenzebach:2014fha,Zhang:2020xub}. The tetrad vectors can be
expressed as

\begin{eqnarray}
{e_{t}} &=&\frac{{\mathcal{K}{(r,\theta )}}}{\sqrt{g(r)}}{\partial _{t}}{%
|_{x_{O}^{\mu }},}  \label{et} \\
&&  \notag \\
{e_{r}} &=&-\mathcal{K}{(r,\theta )}\sqrt{g(r)}{\partial _{r}}{|_{x_{O}^{\mu
}},}  \label{er} \\
&&  \notag \\
{e_{\theta }} &=&\frac{{\mathcal{K}{(r,\theta )}\sqrt{X(\theta )}}}{r}{%
\partial _{\theta }}{|_{{x_{O}^{\mu }}},}  \label{ethe} \\
&&  \notag \\
{e_{\varphi }} &=&-\frac{{\mathcal{K}{(r,\theta )K}}}{{r\sin \theta \sqrt{%
X(\theta )}}}{\partial _{\varphi }}{|_{{x_{O}^{\mu }}}.}  \label{ephi}
\end{eqnarray}

This particular tetrad corresponds to the zero angular momentum observer
(ZAMO) frame, since it satisfies the orthogonality condition $e_{t}\cdot
\partial _{\varphi }=0$ \cite{Grenzebach:2015oea,Sui:2023rfh}. The ZAMO
reference frame is especially useful for describing photon trajectories, as
it provides a locally non-rotating viewpoint for an observer situated in the
curved spacetime surrounding the black hole.

Next, consider the trajectory of a photon, parametrized using the \textit{%
Mino} parameter $\sigma$. The tangent vector to the null geodesic can be
written in two equivalent representations. In coordinate form, it takes the
expression \cite{Frost:2020zcy,Frost:2021bvt}

\begin{eqnarray}
\dot{\rho} &=& \frac{dt}{d\sigma}\partial_{t}+\frac{dr}{d\sigma}\partial_{r}+%
\frac{d\theta}{d\sigma}\partial_{\theta}+\frac{d\varphi}{d\sigma}%
\partial_{\varphi},  \label{eta1}
\end{eqnarray}

whereas in the tetrad frame of the observer, it can be expanded as

\begin{eqnarray}
\dot{\rho} &=& \lambda\Big(-e_{t}+(\cos\alpha) e_{r}+(\sin\alpha\cos\beta)
e_{\theta}+(\sin\alpha\sin\beta) e_{\varphi}\Big).  \label{eta2}
\end{eqnarray}

Here, the parameters $\alpha$ and $\beta$ serve as the celestial coordinates
of the photon as seen by the observer. Specifically, $\alpha$ represents the
angular displacement analogous to latitude, while $\beta$ corresponds to the
longitude angle on the observer's celestial sphere. The coefficient $\lambda$
denotes a normalization constant defined through the relation

\begin{equation}
\lambda =g\left( \dot{\rho},e_{t}\right)
\end{equation}

is introduced. Without the loss of generosity, without any loss of
generality, one may choose $\lambda =-\frac{{r{o^{2}}}}{{\mathcal{K}{{%
(r,\theta )}^{2}}}}$ \cite{Frost:2020zcy,Frost:2021bvt}. By equating the two
expressions for $\dot{\rho}$ in Eqs.~(\ref{eta1}) and (\ref{eta2}), and
substituting the tetrad definitions given in Eqs. (\ref{et})-(\ref{ephi}),
one obtains explicit expressions for the conserved quantities of motion.
These are the photon's energy $E$, angular momentum $L$, and Carter-like
constant $Q$, which are found to be 
\begin{eqnarray}
E &=&\frac{\sqrt{g({r_{O}})}}{{\mathcal{K}{({r_{O}},{\theta _{O}})}}},
\label{EE} \\
&&  \notag \\
L &=&\frac{{{r_{O}}\sqrt{X({\theta _{O}})}\sin {\theta _{O}}\sin \alpha \sin
\beta }}{{\mathcal{K}{({r_{O}},{\theta _{O}})K}}},  \label{LL} \\
&&  \notag \\
Q &=&\frac{{r_{O}^{2}{{\sin }^{2}}\alpha }}{{\mathcal{K}({r_{O}},{\theta _{O}%
})}^{2}}.  \label{QQ}
\end{eqnarray}

These relations provide the direct link between the observer's celestial
coordinates $(\alpha,\beta)$ and the constants of motion that govern photon
geodesics. They will play a central role in the characterization of the
black hole shadow, as they allow us to project the null geodesics onto the
observer's sky and thereby determine the shadow boundary.

\subsection{Angular Shadow and Shadow radius}

We aim to explore how the acceleration, ModMax NLED, and $F(R)$ gravity
parameters influence the angular and azimuthal observational angles, $\alpha 
$ and $\beta $. By combining Eq.~(\ref{LL}) with Eq.~(\ref{QQ}), we obtain
the following relation, hereafter referred to as 
\begin{align}
\sin \alpha =& \delta \frac{\sqrt{g({r_{O}})}}{{{r_{O}}}}, \\
&  \notag \\
\sin \beta =& \zeta \frac{K}{{\sqrt{X({\theta _{O}})}\sin {\theta _{O}}}}.
\end{align}

Furthermore, by substituting the constant $\eta $ from the expressions of
the photon sphere radius in Eqs.~(\ref{photonic1})--(\ref{photonic2}), which
describe the conditions for photon orbits, we arrive at a second relation
governing the angular shadow, denoted as 
\begin{align}
\alpha _{ph}=& \sin ^{-1}\left( \frac{{{r_{ph}}}}{{{r_{O}}}}\sqrt{\frac{{g({%
r_{O}})}}{{g({r_{ph}})}}}\right)  \notag \\
=& \sin ^{-1}\left( \sqrt{\frac{{r_{ph}}^{2}\left( \left( 1-A^{2}{r_{O}}%
^{2}\right) \left( \frac{\frac{q^{2}e^{-\gamma }}{1+f_{R_{0}}}-2m_{0}{r_{O}}%
}{{r_{O}}^{2}}+1\right) -\frac{{R_{0}}{r_{O}}^{2}}{12}\right) }{{r_{O}}%
^{2}\left( \left( 1-A^{2}{r_{ph}}^{2}\right) \left( \frac{\frac{%
q^{2}e^{-\gamma }}{1+f_{R_{0}}}-2m_{0}{r_{ph}}}{{r_{ph}}^{2}}+1\right) -%
\frac{{R_{0}}{r_{ph}}^{2}}{12}\right) }}\right) ,  \label{alpha} \\
&  \notag \\
& {\beta _{ph}}=\sin ^{-1}\left( \sqrt{\frac{{X({\theta _{ph}})}}{{X({\theta
_{O}})}}}\frac{{\sin ({\theta _{ph}})}}{{\sin ({\theta _{O}})}}\right) 
\notag \\
& =\sin ^{-1}\left( \frac{{\sin ({\theta _{ph}})}}{{\sin ({\theta _{O}})}}%
\sqrt{\frac{\frac{q^{2}e^{-\gamma }A^{2}\cos ^{2}(\theta _{p}h)}{1+f_{R_{0}}}%
+2m_{0}A\cos (\theta _{ph})+1}{\frac{q^{2}e^{-\gamma }A^{2}\cos ^{2}(\theta
_{O})}{1+f_{R_{0}}}+2m_{0}A\cos (\theta _{O})+1}}\right) .  \label{beta}
\end{align}%
%
%
%
%

\begin{figure}[tbph]
\centering
\includegraphics[width=0.32\linewidth]{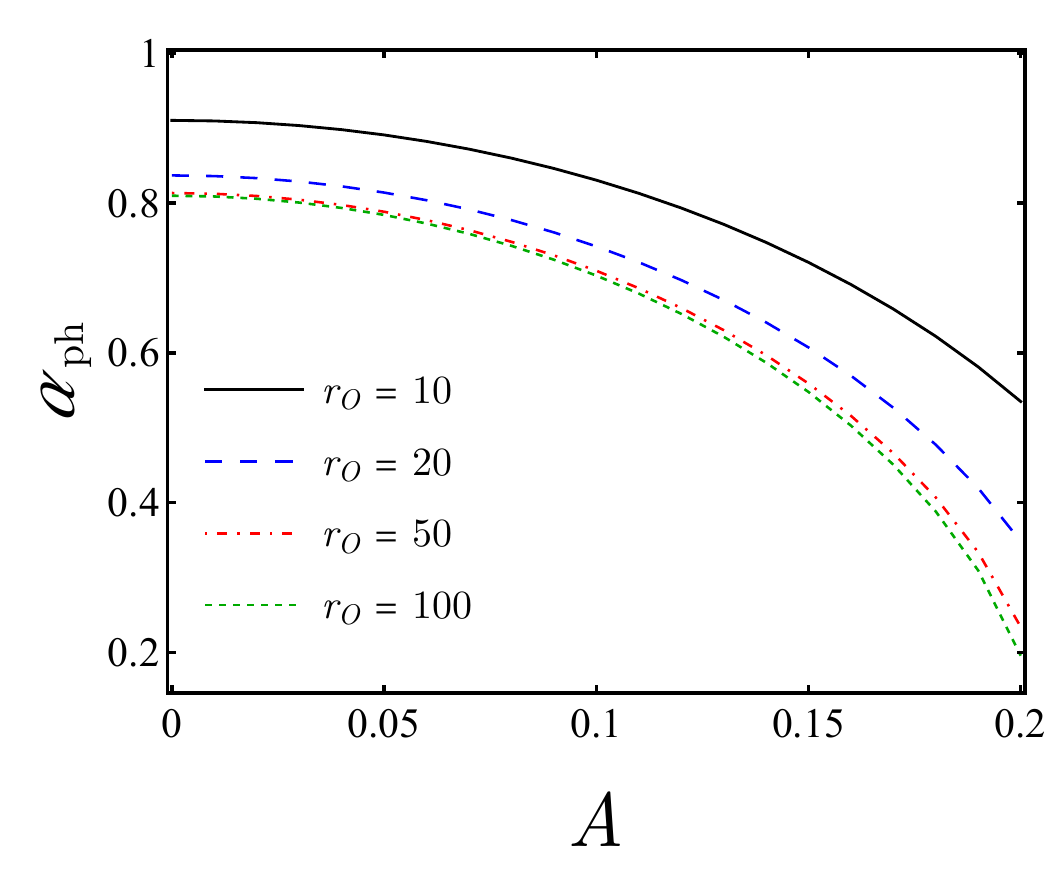} %
\includegraphics[width=0.32\linewidth]{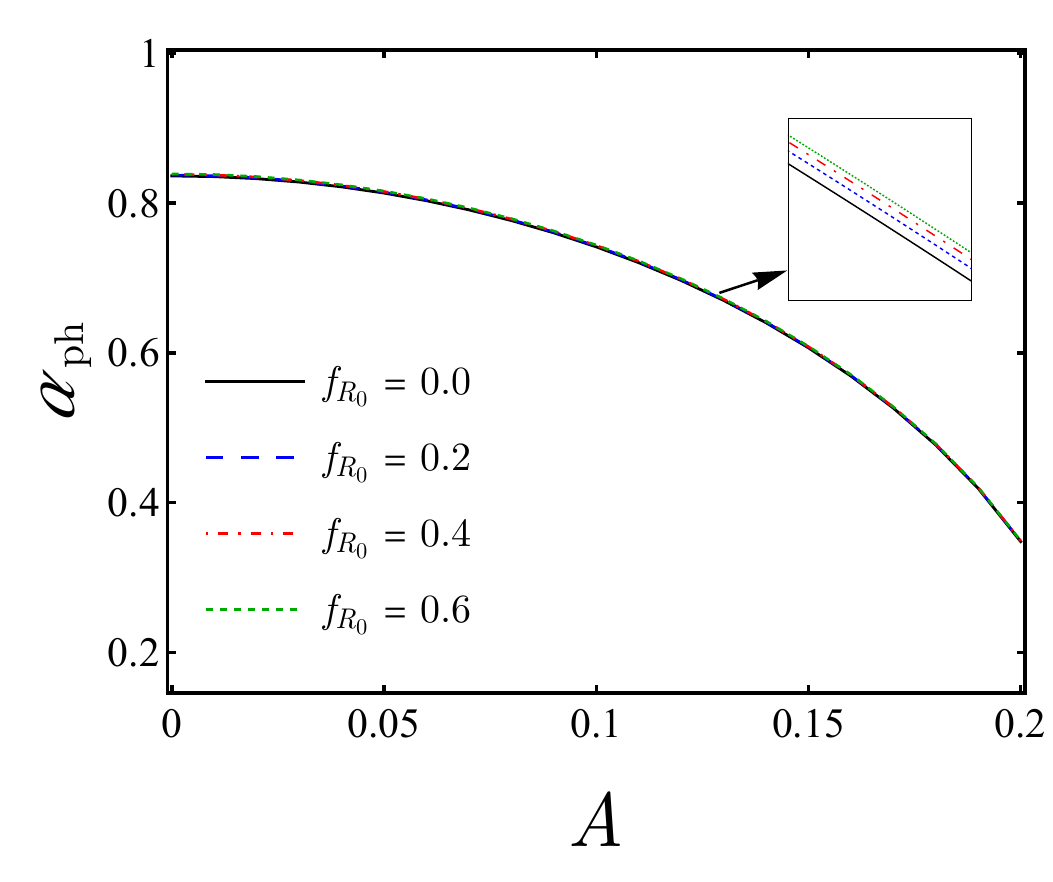} %
\includegraphics[width=0.32\linewidth]{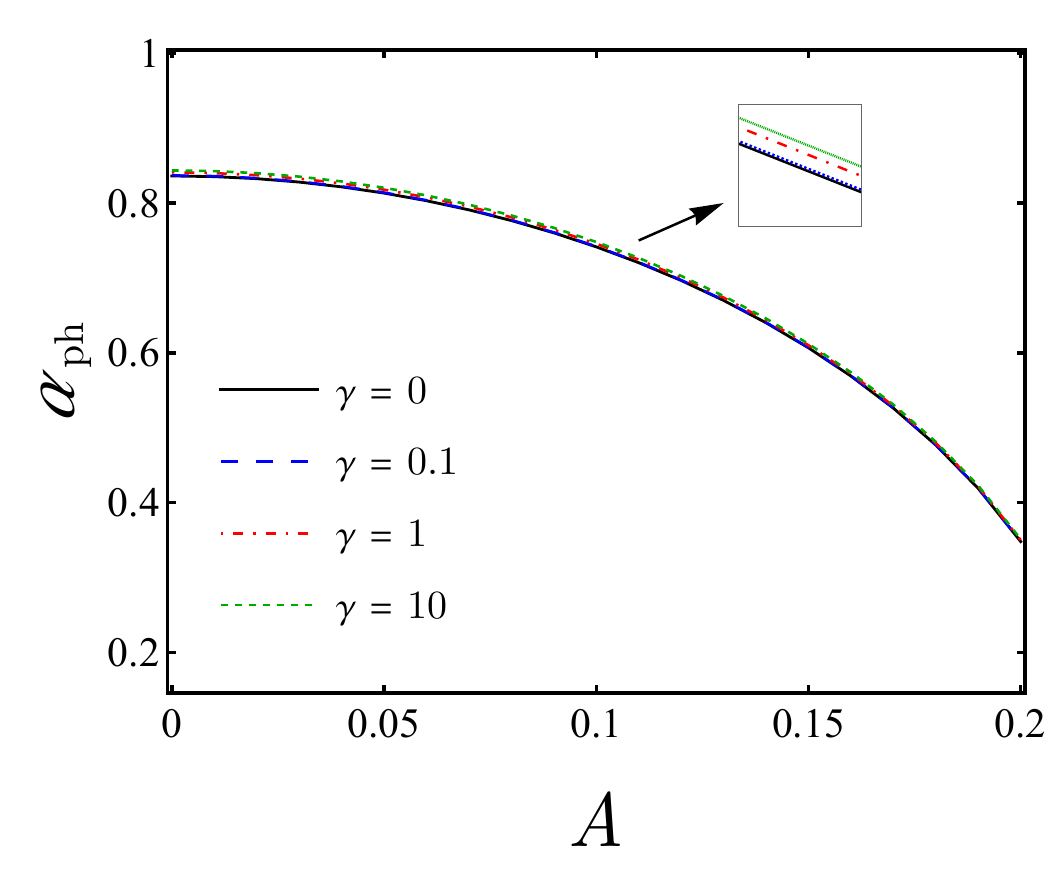} \newline
\caption{Angular shadow radius $\protect\alpha_{ph}$ as a function of the
acceleration parameter $A$ with $m_0 = 1$, $q = 0.3$, and $R_0 = -0.5$. Left
panel: variation with the observer's radial distance $r_O$ at fixed $\protect%
\gamma = f_{R_0} = 0.1$. Middle panel: variation with $f_{R_0}$ for $\protect%
\gamma = 0.3$ and $r_O = 20$. Right panel: variation with $\protect\gamma$
for $f_{R_{0}} = 0.1$ and $r_{O} = 20$. }
\label{fig:AngularShadow}
\end{figure}

Figure \ref{fig:AngularShadow} illustrates how the acceleration parameter $A$
affects the angular shadow under different conditions. In the left panel,
varying the observer's distance $r_{O}$ shows decreasing effects on the
angular shadow at fixed value of acceleration. The role of $A$ is a
consistent reduction in the shadow radius $\alpha _{ph}$. The middle panel
reveals that increasing the $F(R)$ gravity parameter $f_{R_{0}}$ slightly
enlarges the shadow, indicating that higher-order curvature corrections tend
to expand the observable shadow relative to GR. Moreover, the right panel
shows that although the ModMax parameter has a subtle impact, larger values
of the ModMax parameter $\gamma $ makes the shadow radius larger. This
interplay suggests that black hole shadows may encode observable imprints of
both modified gravity and NLED corrections. The resulting equations for $%
\alpha _{ph}$ and $\beta _{ph}$ in Eq. \eqref{alpha} and \eqref{beta} show
that these quantities are independent of each other. Consequently, the
shadow radius takes a circular form. This outcome is somewhat unexpected,
since one would generally anticipate deviations from circularity in the case
of accelerating black holes within a spherically antisymmetric $C-$metric
spacetime. Previous studies \cite%
{Griffiths:2005qp,Grenzebach:2014fha,Grenzebach:2015oea,Heidari:2025llu},
have likewise demonstrated that the shadows of accelerating $C-$metric black
holes remain circular and independent of the observer's inclination angle $%
\theta _{O}$. Thus, one may regard this as a limitation of $C-$metrics, as
their shadows cannot be distinguished from those produced by spherically
symmetric black hole geometries. For visualizing the shadow, we employ a
stereographic projection that maps the celestial sphere of the observer onto
a two-dimensional plane. On this plane, we introduce Cartesian coordinates
defined in a dimensionless form, following the approach of \cite%
{Perlick:2021aok} 
\begin{eqnarray}
x &=&-2\tan \left( \frac{{{\alpha _{ph}}}}{2}\right) \sin \left( {\beta _{ph}%
}\right) , \\
&&  \notag \\
y &=&-2\tan \left( \frac{{{\alpha _{ph}}}}{2}\right) \cos \left( {\beta _{ph}%
}\right) .
\end{eqnarray}%
%
%
%
%
%
%
%
%
%
%
%
%

The shadow radius is then determined through the expression provided in 
\begin{equation}
R_\text{sh}=\sqrt{x^2+y^2}.
\end{equation}

\begin{figure}[tbph]
\centering
\includegraphics[width=0.32\linewidth]{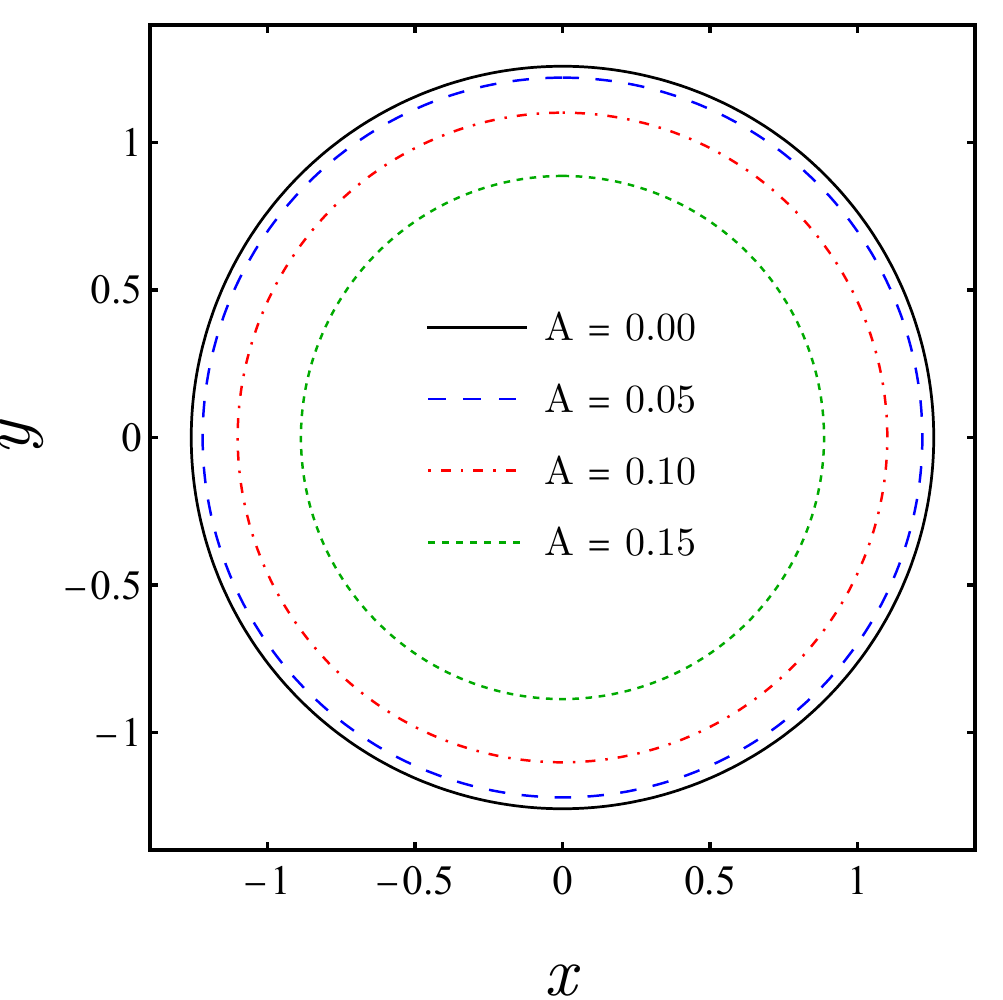} \includegraphics[width=0.32%
\linewidth]{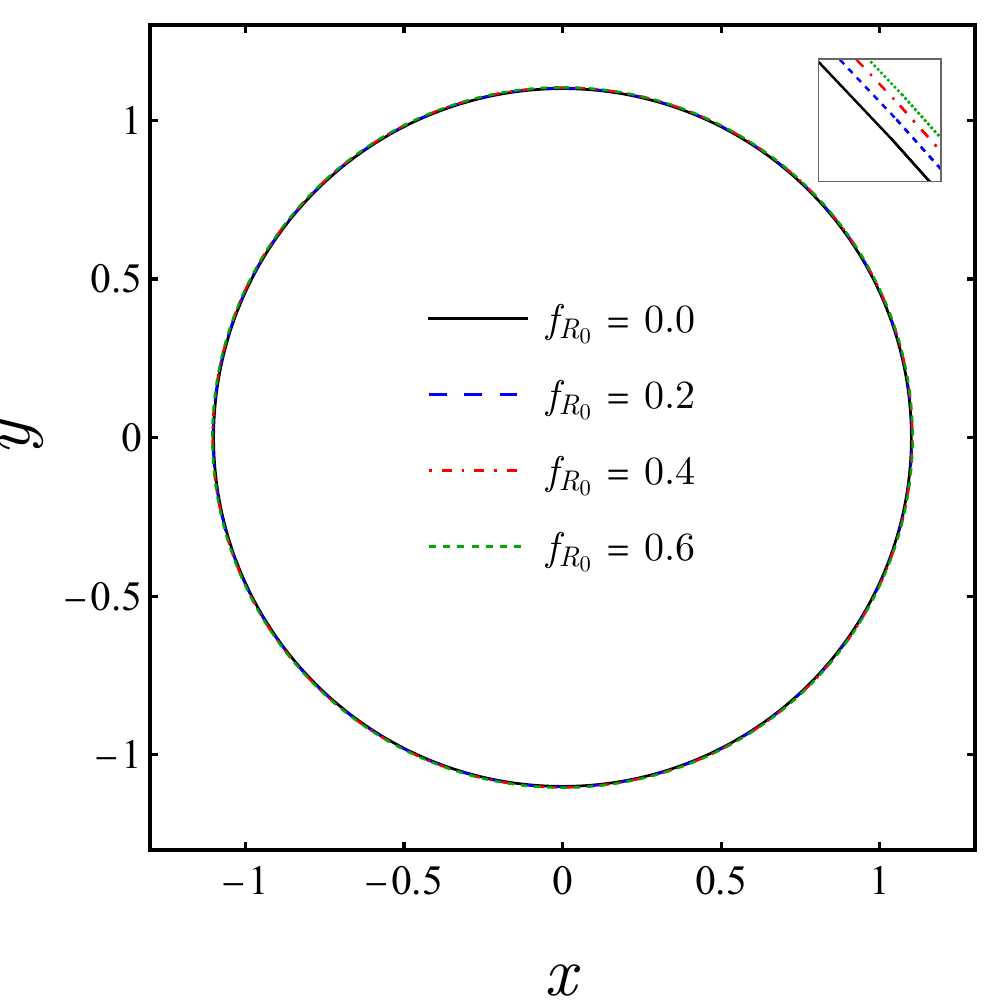} \includegraphics[width=0.32\linewidth]{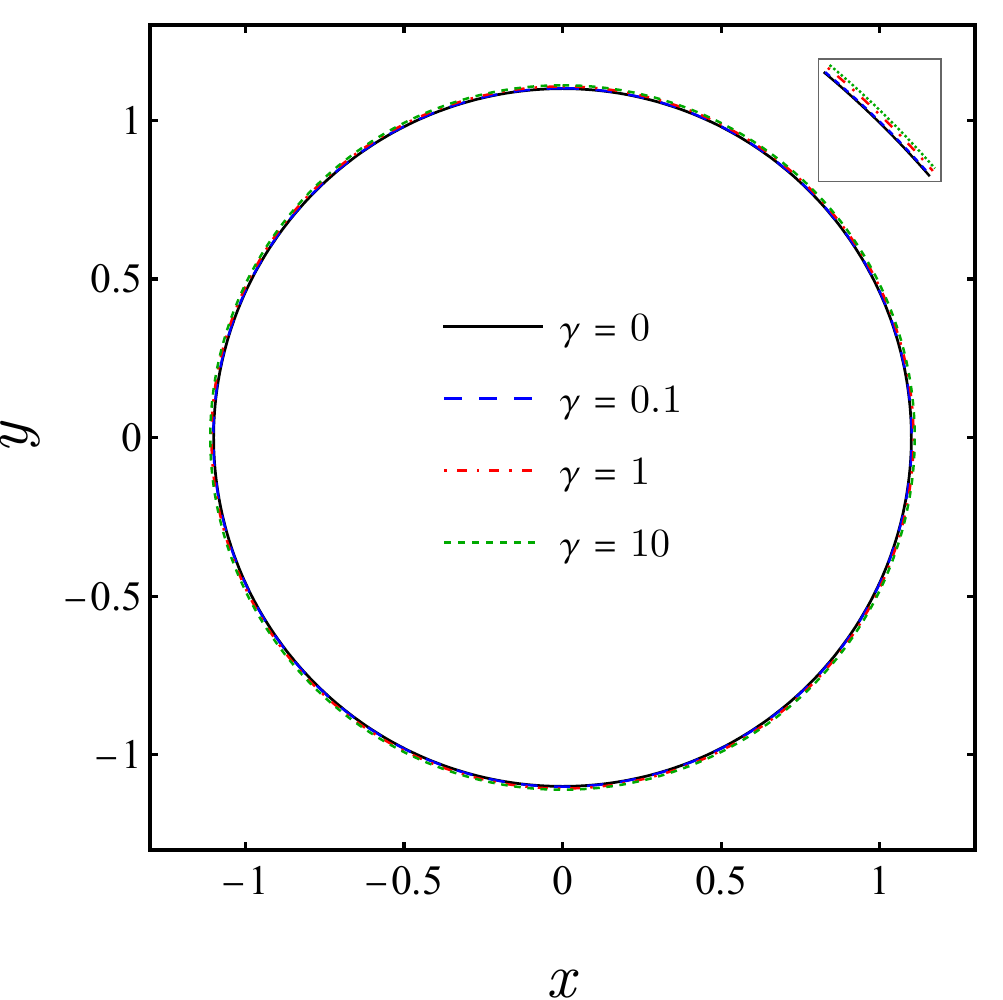} 
\newline
\caption{Dependence of the shadow radius on the model parameters. The shadow
radius is calculated with fixed parameters $m_0 = 1 $, $q = 0.3 $, $R_0 =
0.5 $, and $r_O = 20 $. The left panel, variation with the acceleration
parameter $A $ (for $f_{R_0} = 0.1 $, $\protect\gamma = 0.1 $). The middile
panel, variation with the parameter $f_{R_0} $ (for $A = 0.1 $, $\protect%
\gamma = 0.1 $). The right panel, variation with the parameter $\protect%
\gamma $ (for $A = 0.1 $, $f_{R_0} = 0.1 $).}
\label{fig:Shadow}
\end{figure}

Finally, Fig.~\ref{fig:Shadow} illustrates the effects of the acceleration,
ModMax parameter, and $F(R)$ gravity modifications on the shadow radius of
the accelerating black hole. These results highlight the role of NLED and
modified gravity corrections in shaping the observable properties of black
hole shadows.

The results of our parametric analysis, depicted in Fig.~\ref{fig:Shadow},
reveal the distinct and competing influences of the acceleration parameter $%
A $, the modified gravity parameter $f_{R_0}$, and the ModMax parameter $%
\gamma $ on the black hole shadow. The left panel demonstrates that the
acceleration parameter $A$ induces a pronounced inward deformation, breaking
circular symmetry. This effect, which dominates over other geometric
factors, consistently reduces the effective shadow radius and is a
characteristic signature of the black hole. The middle panel indicates that
the $F(R)$ gravity parameter $f_{R_0}$ counteracts this reduction; larger
values systematically enlarge the shadow, suggesting that higher-order
curvature corrections in modified gravity expand the photon capture region
relative to GR. Likewise, the right panel shows that the ModMax parameter $%
\gamma$ increases the shadow radius. Overall, these results highlight that
the shadow's morphology and size are not determined by a single effect but
by a balance between them; acceleration reduces it, modified gravity and
NLED tend to enlarge it. Consequently, the black hole shadow serves as a
potent observational discriminant for these competing high-energy and
gravitational corrections. 

\section{Conclusions}

We combined $F(R)$ gravity with the ModMax NELD theory, resulting in what we
refer to as $F(R)-$ModMax theory. From this, we derived the equations of
motion for the theory. By applying the $C-$metric within the framework of $%
F(R)-$ModMax theory, we obtained an exact black hole solution, which we
named the accelerating ModMax black hole in $F(R)$ gravity. We then
calculated the Kretschmann scalar to investigate how various parameters
influenced this quantity. Our analysis of the asymptotic behavior of the
spacetime revealed the following:

i) The spacetime does not exhibit asymptotic (A)dS behavior.

ii) The ModMax parameter ($\gamma$) does not influence the asymptotic
behavior of the spacetime.

We investigated how various parameters of $F(R)-$ModMax theory affect the
event horizon. Our findings are summarized as follows:

i) As the electric charge $q$ increased, the number of roots decreased. This
indicates that a highly charged black hole in $F(R)-$ModMax theory lacks an
event horizon, leading to the formation of a naked singularity.

ii) Increasing the parameter $\gamma$ resulted in an increase in both the
number of roots and the radius of the event horizon. Specifically, a black
hole with a large $\gamma$ exhibited two roots.

iii) An increase in $f_{R_{0}}$ was associated with a rise in both the
number of roots and the radius of black holes.

iv) Conversely, when $\left\vert R_{0}\right\vert$ increased, both the
number of roots and the radius of the black hole decreased.

By examining the behavior of $q$ and $\gamma $ on the event horizon of
accelerating ModMax black holes in $F(R)$ gravity, we discovered that the
ModMax parameter behaves oppositely to the electric charge.

We extracted the Hawking temperature of accelerating ModMax black holes in $%
F(R)$ gravity and examined how various parameters influence this quantity.
Our findings revealed the following:

i) There exists a critical value for the acceleration parameter, $A_{\text{%
crit}} = \sqrt{\frac{-R_{0}}{12}}$, which results in two distinct behaviors
for the temperature of large black holes. Specifically, the temperature of
large black holes is negative when $A > A_{\text{crit}}$ and positive when $%
A < A_{\text{crit}}$.

ii) The temperature has two roots and one divergence point, remaining
positive in two regions: $r_{+_{root_{1}}} < r_{+} < r_{+_{div}}$ and $r_{+}
> r_{+_{root_{2}}}$ (when $A < A_{\text{crit}}$).

iii) There is a critical value for the ModMax parameter, $\gamma_{\text{crit}%
}$. For $\gamma > \gamma_{\text{crit}}$, the Hawking temperature of small
black holes is positive; however, for $\gamma < \gamma_{\text{crit}}$, these
small black holes cannot be considered physical systems.

iv) The small root ($r_{+_{root_{1}}}$) of the temperature is dependent on $%
f_{R_{0}}$. Specifically, as $f_{R_{0}}$ increases, the physical area
expands because the small root of $T$ decreases with increasing $f_{R_{0}}$.

We also calculated the entropy of accelerating ModMax black holes in $F(R)$
gravity and examined how the parameters of acceleration and $F(R)$ gravity
affect this quantity. Our findings indicate that the asymptotic behavior of
the entropy can be positive when $A < \sqrt{\frac{-R_{0}}{12}}$, which
requires the scalar curvature of spacetime to be negative (i.e., $R_{0} < 0$%
). Conversely, we identified a divergence point for the entropy when the
acceleration parameter takes on large values. Notably, this divergence point
shifts to smaller radii as $A$ increases. In other words, for very large
values of $A$, the entropy becomes negative at all radii. To avoid this
divergence point in the entropy, we need to consider smaller values for $A$.

To study the local stability of accelerating ModMax black holes, we examined
their heat capacity. Our findings revealed four roots and two divergence
points for the heat capacity. Additionally, we analyzed the temperature,
entropy, and heat capacity of the accelerating ModMax black holes to
identify their physical and local stability regions. We discovered seven
distinct regions, two of which satisfied the conditions for physical and
thermal stability. Specifically, the accelerating ModMax black holes in $%
F(R) $ gravity can simultaneously meet the physical and local stability
conditions when their radius falls within the second region (between the
first root of temperature and the first divergence point of heat capacity,
i.e., $r_{T_{1}=0}<r_{+}<$ $r_{C_{1}=\infty}$) and the fourth region
(between the second divergence point and the second root of the heat
capacity, i.e., $r_{C_{2}=\infty}<r_{+}<r_{C_{2}=0}$). Furthermore, we
investigated the effect of $A$ on these regions and found that:

i) Increasing $A$ decreases the physical and stable areas (the second and
fourth regions).

ii) A phase transition occurs between the second region (Phase 1) and the
fourth region (Phase 2), which disappears when $A$ takes on very large
values.

We also examined how the parameters of $F(R)$ gravity affect these regions.
Our analysis showed that the second region ($r_{T_{1}=0} < r_{+} <
r_{C_{1}=\infty}$) decreased as $f_{R_{0}}$ increased.

In conclusion, our study demonstrated that the shadow of a black hole within
the $F(R)-$ModMax framework offers a valuable phenomenological insight. The
observable shadow radius and its deformation are influenced by multiple
factors rather than a single parameter. Specifically, acceleration primarily
reduces the shadow size, while modifications from $F(R)$ gravity and the
ModMax NLED tend to increase it. This complex interplay creates an
observationally testable signature that could help distinguish the effects
of acceleration from those of modified gravity or NELD in future
high-resolution astrophysical observations.

A notable finding of this work is that the obtained solutions in the electric sector exhibit a formal equivalence to the $f(R)-$Maxwell framework through the identification of an effective charge $Q=qe^{-\gamma /2}$. Rather than rendering the results trivial, this equivalence elucidates the specific contribution of the ModMax parameter to the spacetime structure. It demonstrates that the impact of the non-linear parameter $\gamma$ can be effectively captured by a rescaling of the source term, a feature that persists as long as magnetic components are neglected, thereby simplifying the analysis of the black hole's thermodynamic and geometric properties.

\acknowledgements{B. Eslam Panah thanks the University of Mazandaran. This work is based upon research funded by Iran National Science Foundation (INSF) under project No.4035285. A.~R. would like to express his gratitude to Silesian University in Opava, Czech Republic, for their financial support. A.~R. is very grateful for the hospitality of the University of Valencia (Spain), Valencia Polytechnic University (Spain) and the Complutense University of Madrid (Spain). The creation of this article was supported by the grant program Vouchers for Universities in the Moravian-Silesian Region (registration number CZ.10.03.01/00/23\_042/00003901119). This article is based upon work from COST Action FuSe, CA24101, supported by COST (European Cooperation in Science and Technology).}

\end{document}